\documentclass[pre,aps,twocolumn,showpacs,superscriptaddress]{revtex4}
\usepackage{bm}
\usepackage{epsfig}
\usepackage{amsmath}
\usepackage{amssymb}

\usepackage[usenames]{color}
\definecolor{Blue}{rgb}{0,0.0,1.0}
\definecolor{Red}{rgb}{1,0.0,0.0}
\definecolor{Grey}{rgb}{0.6,0.6,0.6}

\newcommand{\ee}{\mathrm{e}}

\begin{document}

\title{Random pinning in glassy spin models with plaquette interactions}
\author{Robert L. Jack}

\affiliation{Department of Physics, University of Bath, Bath, BA2 7AY, 
United Kingdom}

\author{Ludovic Berthier}

\affiliation{Laboratoire Charles Coulomb, UMR 5221, CNRS and Universit\'e
Montpellier 2, Montpellier, France}

\begin{abstract}
We use a random pinning procedure to study amorphous 
order in two glassy spin models.
On increasing the concentration of pinned spins at constant
temperature, we find a sharp crossover (but no thermodynamic
phase transition)
from bulk relaxation to localisation
in a single state. At low temperatures, both models exhibit scaling behaviour.
We discuss the growing length and time scales associated with amorphous order,
and the fraction of pinned spins 
required to localize the system in a  single state.
These results, obtained for finite dimensional interacting models, 
provide a theoretical scenario for the effect of random pinning 
that differs qualitatively from previous approaches based either 
on mean-field, mode-coupling, 
or renormalization group treatments.
\end{abstract}



\maketitle

\newcommand{\Cs}{g_\mathrm{s}}  
\newcommand{\Nf}{\hat{N}_\mathrm{f}}  
\newcommand{\Nm}{\hat{N}_\mathrm{m}}   
\newcommand{\QQ}{\hat{Q}}  
\newcommand{\gij}{g_{4,ij}}  
\newcommand{\Xij}{X_{ij}}    
\newcommand{\df}{d_\mathrm{f}} 

\newcommand{\Zf}{Z_\mathrm{f}} 

\section{Introduction}
\label{sec:intro}

Supercooled liquids and glasses have very large relaxation
times and complex relaxation mechanisms, but their 
structures appear disordered and 
unremarkable~\cite{ediger96,deb-still01,rmp_bb}.  
This combination is surprising and rather mysterious, 
especially because
several recent studies~\cite{bouchaud2004,montanari2006,kurchan2011} 
indicate that if a 
relaxation time increases sufficiently rapidly on cooling then this must 
be accompanied by the development of some kind of structural order. 
The idea of growing amorphous order is 
that the diversity of amorphous states, as 
quantified by the configurational part of the entropy, decreases
at low temperatures. This leads to increasing
static correlation lengthscales, which can be 
measured using point-to-set 
correlations~\cite{bouchaud2004,montanari2006,cavagna2007,cavagna2008,kurchan2011,kob2011,bookdh}. 
If these length scales are large,
the system may be localized into a single 
amorphous state by fixing the positions of a 
small fraction of the particles~\cite{berthier-kob-pts,cammarota2011}. 
This localization occurs at temperatures above the glass
transition temperature, so the pinning procedure bypasses the challenging
task of thermalizing 
systems at very low temperatures.

Here, we investigate amorphous order in two finite dimensional
interacting spin systems. These
are the square plaquette model (SPM)~\cite{lipowski1997} and the triangular 
plaquette model (TPM)~\cite{newman1999}. They both
have growing relaxation times at low temperatures,
but two-point thermodynamic correlation functions 
do not indicate the presence of any  
growing length scale~\cite{garrahan2002-plaq}, and the models 
do not have phase transitions at any finite temperature. Additionally,  
structural relaxation at low temperatures exhibits strong dynamical 
heterogeneity and growing dynamic 
lengthscales~\cite{plaq-static2005,plaq-caging2005}. These features
mimic those of supercooled liquids, but the models are numerically and 
analytically more tractable than off-lattice particle 
systems. This makes them useful models for studying 
generic features of glassy 
systems~\cite{buhot2002,plaq-static2005,plaq-caging2005,plaq-fdt2006}.

Several methods have been proposed for characterizing amorphous order. 
A prominent recent example
is to fix particles everywhere except within a small cavity, and then
to study the motion of particles within this 
cavity~\cite{bouchaud2004,cavagna2007,cavagna2008,sausset2011,berthier-kob-pts}.  
Other possible measurements of amorphous order involve freezing spins in a 
different geometrical
arrangement~\cite{kob2011,berthier-kob-pts}, or
direct measurement of the diversity of amorphous 
packings~\cite{kurchan2011,sausset2011}. Here we 
follow~\cite{kim2003-pin,cammarota2011,berthier-kob-pts,procaccia2011-pin,tarjus2011pin} and
fix a fraction $f$ of particles (or spins), distributed randomly through
the system in the positions (or orientations) that are representative of
thermal equilibrium.  
In supercooled liquids, a thermodynamic phase
transition on varying $f$ was recently predicted~\cite{cammarota2011}, 
based on an analysis within 
random first-order transition (RFOT) 
theory~\cite{kirk1987,ktw1989}. 
Another recent analysis based on mode-coupling 
theory predicts a dynamic singularity on 
varying $f$~\cite{krakoviack2011-pin}.
Numerical studies~\cite{kim2003-pin,berthier-kob-pts,procaccia2011-pin,tarjus2011pin} 
have mainly addressed the behaviour at
low to moderate pinning fraction $f$, leaving open the 
question of the existence of phase transitions at larger $f$. 

Here, we explore the effect
of random pinning in the plaquette models, concentrating on
the behaviour at the relatively large
pinning fractions where phase transitions are predicted to occur
in supercooled liquids. 
We show that
the plaquette models do not exhibit
thermodynamic phase transitions on 
varying $f$ at any finite temperature. However, we do find a
well-defined crossover from `bulk' behaviour (where the system explores a 
large number of configurations),
to `frozen' behaviour (where the system remains trapped in a region of 
configuration space) occurring at some finite fraction $f^*$.
As the temperature is reduced, the crossover 
occurs at an increasingly small value of 
$f^*$, and becomes increasingly sharp.

From previous studies of point-to-set correlations in closed cavities for plaquette models~\cite{plaq-caging2005}, 
it is known that these models do have growing amorphous order at low temperatures, but
their behaviour is not consistent with RFOT. 
The results we present here therefore illustrate a 
theoretical scenario, alternative to mode-coupling and 
RFOT treatments, for understanding the behaviour of 
model supercooled liquids with randomly pinned particles.

After defining our models in Sec.~\ref{sec:models}, we show numerical results 
for this crossover 
in Sec.~\ref{sec:pinning}.  In Sec.~\ref{sec:lengths} we investigate the 
length scales that characterize
amorphous order in the system, including their scaling with temperature.  
In Sec.~\ref{sec:weak-strong} we
use analytic arguments in the SPM to study the small-$f$ behaviour, and the 
behaviour near the crossover.
Finally in Sec.~\ref{sec:conc} we discuss the interpretation of our results, 
including their relation to
previous theoretical analysis.

\section{Plaquette models}
\label{sec:models}

\subsection{Definitions}

The plaquette models consist of Ising spins, $s_i=\pm 1$,
on a lattice, interacting through either three- or four-body terms.  
They evolve dynamically through single spin-flips with 
Metropolis rates, $w_i=\min(1,\ee^{-\beta \Delta E_i})$, where
$\Delta E_i$ is the change in energy of the system on flipping spin $i$
and $\beta\equiv 1/T$ is the inverse temperature.

The energy of the square plaquette model (SPM) is 
\begin{equation}
E = -\frac12 \sum_\square s_1 s_2 s_3 s_4,
\end{equation} 
where the sum runs over plaquettes of the square
lattice~\cite{lipowski1997,garrahan2002-plaq}.  
It is useful to define $c\equiv\ee^{-\beta}$, since length
and time scales typically show power law scaling with $c$ 
in the low temperature limit ($c\to0$).
The model has a dual representation in terms of `defect' variables, 
$n_p=(1-s_1 s_2 s_3 s_4)/2$, which
are associated with the plaquettes $p$ of the square lattice.  
In terms of these defect variables, the thermodynamic
properties of the system are those of a non-interacting lattice 
gas with $\langle n_p \rangle = (1+\ee^{\beta})^{-1}$ 
approximately equal to $c$ at low temperatures.  
We use square systems with periodic boundaries.

The energy of the triangular plaquette model (TPM) is 
\begin{equation}
E = - \frac12 \sum_\nabla s_1 s_2 s_3,
\end{equation} 
where the sum runs over downward pointing plaquettes of the hexagonal
lattice~\cite{newman1999}.  In this model, the defect variables 
are associated with these plaquettes, $n_p=(1-s_1 s_2 s_3)/2$. We
use rhombus-shaped systems with periodic boundaries.  We always 
take the linear system size to be a power of 2, which
minimizes finite size effects.  

In the defect representation, both the SPM and the TPM  
have strong similarities with 
kinetically constrained models~\cite{ritort2003}.  
In both systems, the thermodynamic
free energy per spin (in the limit of 
large systems) is simply $-T \log [2\cosh(1/2T)]$. This is a
smooth and unremarkable function of temperature
down to $T=0$. However, 
the dynamical properties of the models, to be 
described shortly, are not trivial. 
This combination of simple thermodynamic properties 
and complex dynamics
is consistent with the dynamic facilitation 
picture of the glass transition~\cite{gc2010}. 

\subsection{Square plaquette model at $f=0$}

Before considering the effect of pinning a fraction of spins,
we first review some relevant results for the SPM with $f=0$,  
see Refs.~\cite{garrahan2002-plaq,plaq-static2005} for
more details.
Symmetry under global spin reversal implies that $\langle s_i\rangle=0$, 
and all non-trivial
two-point and three-point correlation functions also vanish.
The simplest non-trivial correlation functions involve four spins. 
For spins $a,b,c,d$ lying on the four 
vertices of an $x \times y$ rectangle then
\begin{equation}
\Cs(x,y) \equiv \langle s_a s_b s_c s_d \rangle_0 = \tanh(\beta/2)^{xy}.
\label{equ:fourspin}
\end{equation}
All other non-trivial four-point correlations vanish.

The function $\Cs(x,y)$ has scaling behaviour at low temperatures.  
Full details are given in Sec.~\ref{subsec:f0scal} below, but
the essential points are that correlations are anisotropic and
the model supports two correlation lengths at low temperatures.  The
shorter length is $\xi_0 \sim c^{-1/2}$, and there is also a longer 
length $\xi \sim c^{-1}$ associated with
the correlations along the lattice axis.  
We note that previous analysis of static correlations 
considered only the shorter of these  
two correlation lengths~\cite{plaq-static2005}. 

Point-to-set correlations were considered in
Ref.~\cite{plaq-caging2005}, for 
the case of an isotropic (square) cavity of linear size $r$.
For $r \lesssim \xi_0 \sim c^{-1/2}$, 
the configurational entropy of the cavity is 
close to zero.  For  $\xi_0 \lesssim  r \lesssim \xi \sim c^{-1}$, the cavity
has strong finite size effects but its configurational entropy is 
non-zero and the spin-spin autocorrelation function decays to zero at
the centre of the cavity. This implies that these larger cavities are no longer frozen in 
a single amorphous state, so we identify
$\xi_0$ as the point-to-set length~\cite{plaq-caging2005}.

Dynamical observables in the square plaquette 
model also show scaling behaviour.  
The relaxation time for spins is $\tau \approx c^{-3} \sim 
\ee^{3/T}$, consistent
with the behaviour of `strong' glass-formers~\cite{buhot2002}.
Energy-energy  
correlations decay have relaxation times longer than $\tau$ at low temperatures, 
although they also have Arrhenius
scaling.
Four-point dynamic correlation functions, also discussed in more
detail below, are anisotropic as well, with correlations being 
strongly localized along the axis of the square lattice, 
and having spatial on-axis extension of range 
$\xi \sim \ee^\beta = c^{-1}$~\cite{plaq-static2005}.

\subsection{Triangular plaquette model at $f=0$}

Static correlations in the TPM were discussed in Ref.~\cite{newman1999} where 
it was shown that $\langle s_i\rangle=0$ and
$\langle s_i s_j\rangle = \delta_{ij}$, as in the SPM.  The relaxation 
time in the TPM  has `fragile' super-arrhenius scaling 
$\tau \sim \ee^{1/(T^2\log 3)}$~\cite{plaq-caging2005},
which arises from a hierarchy of mechanisms whose characteristic 
timescales increase as the logarithm of their associated
lengthscales (see also~\cite{gc2003-pnas,nef2005,keys2011}).  
Static correlation 
lengths, point-to-set correlations, and dynamical
four-point correlations are considered in Ref.~\cite{plaq-caging2005}, 
where it was found that the scaling of these functions
all depend on a unique correlation lengthscale, 
$\xi \sim c^{-1/\df}$, where $\df=\log 3/\log 2\approx 1.585$ is the 
fractal dimension of the Sierpinski triangle.  It was also found 
recently~\cite{sasa2010} that the TPM 
supports unusual phase transitions if it is constrained to have
a fixed non-zero magnetisation, but we do not discuss that case here.

\section{Effect of random pinning}
\label{sec:pinning}

We now turn to our main results, which correspond to the following 
thought-experiment~\cite{cammarota2011,kim2003-pin,kim2011-pin,procaccia2011-pin,berthier-kob-pts,krakoviack2011-pin}.  
We select a reference configuration from a thermally equilibrated 
system and we instantaneously `freeze' (or `pin') at time $t=0$ the 
state of a finite set of spins. Each spin is 
frozen with probability $f$, independently of all other spins.
The spins that are not frozen evolve with Monte-Carlo
(MC) dynamics for some time $t$, after which various measurements 
are performed.

\subsection{Correlation functions and susceptibilities}

To describe these measurements, we first establish our notations and define the correlation functions 
and susceptibilities that we will consider.
The system consists of $V$ spins $s_i=\pm1$, with $i=1, \dots,  V$.  
To describe whether a spin is frozen, we introduce the 
binary variables $f_i$, with $f_i=1$ if spin $i$ is frozen, zero otherwise.  
We use angle brackets $\langle \cdots \rangle$ to denote averages which 
run over the reference configuration, 
the choice of frozen spins,
and the MC dynamics.  
We also use $\langle \cdots \rangle_0$ to denote a `bulk' thermal average,
i.e. in the absence of any pinned particles ($f=0$).

It is convenient to define the autocorrelation of the mobile spins. For 
spin $i$, this is defined as $a_i(t) = (1-f_i) s_i(t) s_i(0)$, and it 
takes values $0,\pm1$.
The number of mobile (unfrozen) spins is $\Nm = \sum_i (1-f_i)$,
which is a fluctuating quantity within our analysis.
We also define
the (extensive) overlap $\QQ(t)=\sum_i a_i(t)$ and the 
(intensive) autocorrelation function
\begin{equation}
C(t) = \frac{ \langle \QQ(t) \rangle }{ \langle \Nm \rangle } = \frac{ 
\langle a_i(t) \rangle }{ 1-f } .
\end{equation}
The four-point susceptibility quantifies the strength of spontaneous 
fluctuations of the overlap, 
\begin{equation}
\chi_4(t) =  \frac{ \langle \delta \QQ(t)^2 \rangle }{ \langle \Nm \rangle }
 = \frac{ \sum_j \gij }{ 1 - f },
\label{equ:chi4}
\end{equation}
and it is related
to the volume integral of the four-point correlation function
defined as
\begin{equation}
\gij =  \langle \delta a_i(t) \delta a_j(t) \rangle.
\end{equation}
We use the notation $\delta O = O - \langle O \rangle$ 
for thermal fluctuations throughout this article.

In addition to the usual 
four-point correlations, we introduce three-point correlations 
and susceptibilities. We measure how the positions of the 
pinned spins are correlated with the states of mobile spins through
the correlation
\begin{equation}
\Xij(t) = \langle \delta a_i(t) \delta f_j \rangle.
\end{equation}
The ratio $(\Xij/f)$ measures the change in the average of $a_i$ if 
spin $j$ is assumed to be frozen. Hence,
\begin{equation}
\frac{\partial C(t)}{\partial f} = \frac{1}{f(1-f)^2} \sum_{j \neq i} \Xij.
\label{equ:xchi}
\end{equation}
To prove this relation, one writes $f=(1+\ee^{\mu})^{-1}$, so that
$\langle a_i(t) \sum_j \delta f_j\rangle = -\frac{\partial}{\partial 
\mu}\langle a_i(t)\rangle$, and
Eq.~(\ref{equ:xchi}) follows.  

It is interesting to note that $\Xij(t)$ is 
an example of a three-point correlation function,
and that $\frac{\partial C(t)}{\partial f}$ is therefore a three-point 
susceptibility, in the spirit of the dynamic functions 
introduced in Refs.~\cite{berthier2007-jcp1,berthier2007-jcp2}.
Following Refs.~\cite{berthier2005-science,berthier2007-jcp1,berthier2007-jcp2} 
we can obtain a lower bound $\Delta \chi_4$ on the 
four-point susceptibility, $\chi_4(t) \geq \Delta \chi_4(t)$,
which involves the three-point susceptibility defined above as
\begin{equation}
\Delta \chi_4(t) = f \left[ (1-f)\frac{\partial C(t)}{\partial f} - 
C(t) \right]^2.
\label{equ:deltachi}
\end{equation} 
This relation is proved in Appendix~\ref{app:pin} where we also show that 
a sufficient condition for saturation of the 
bound [i.e., $\chi_4(t)=\Delta\chi_4(t)$]
is that for a given choice of frozen spins, the autocorrelations of the 
mobile spins are 
independent of each other and depend separately on the individual $f_j$.
When this condition is obeyed, then
the four-point correlation function may be expressed as a convolution 
of $\Xij$ with itself:
\begin{equation}
\gij(t) \approx  \frac1{f(1-f)} \sum_k X_{ik}(t) X_{jk}(t).
\label{equ:g4conv}
\end{equation}
We refer to Appendix~\ref{app:pin} for more details on this 
particular point.

Of particular interest is the limit of large time, $t \to \infty$, 
in which correlation functions become independent
of the dynamical evolution of the system and may be calculated by 
equilibrium statistical mechanics.  In this limit, 
we drop the time argument on our correlation functions, writing, as 
$t\to\infty$:
$C(t) \to q$, $\gij(t) \to \gij$, $\Xij(t) \to \Xij$,
and $\chi_4(t) \to \chi_4$. These limiting quantities 
are given by `static' correlation
functions. A general recipe for calculating them is given in 
Appendix~\ref{app:pin}, which follows a similar analysis 
for off-lattice particle systems~\cite{krak2010}.  
For small $f$, in particular, 
they may be calculated in a series expansion, see 
also Sec.~\ref{sec:weak-strong}.

\subsection{Numerical results}

\begin{figure*}
\includegraphics[width=18cm]{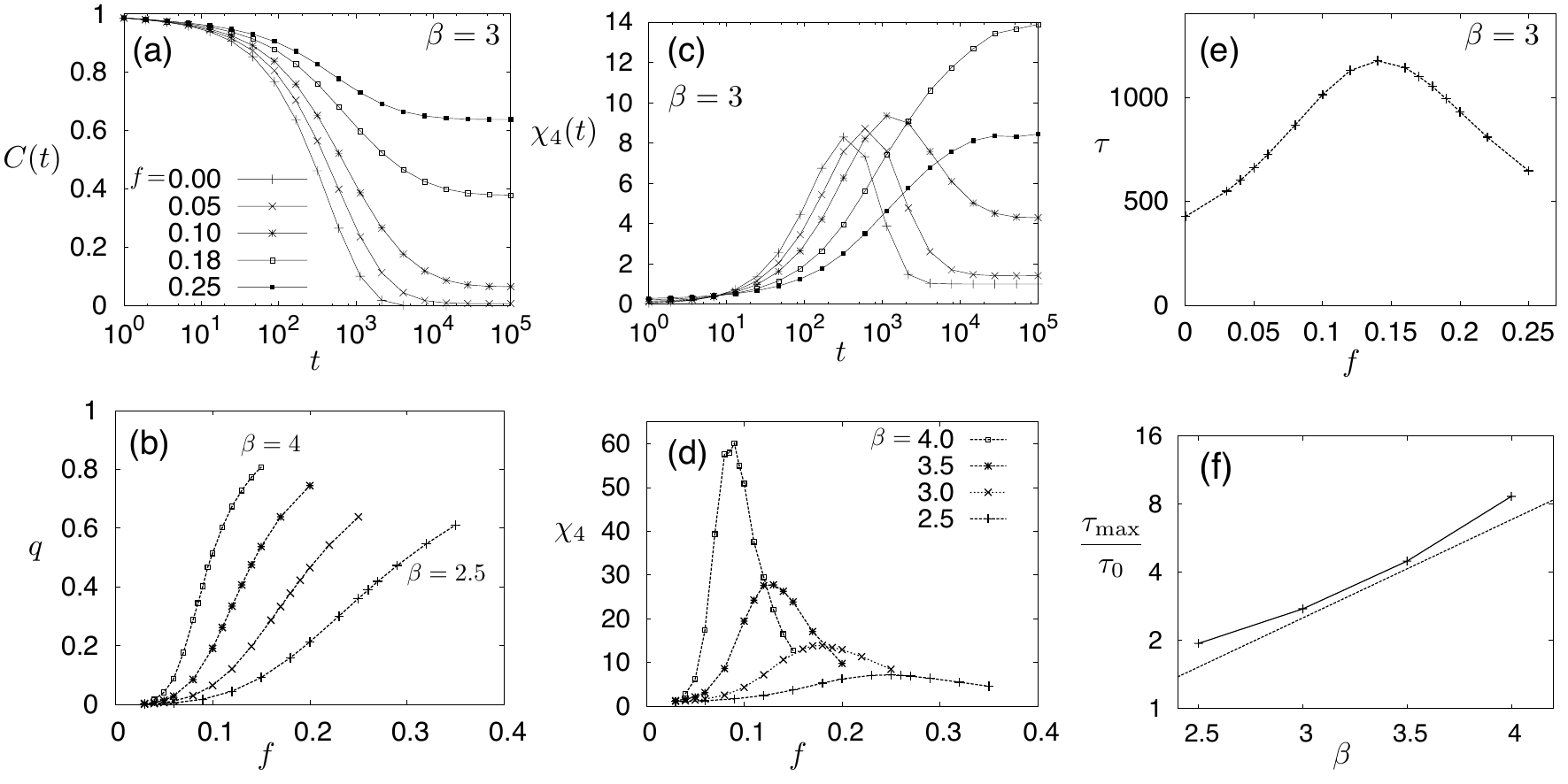}
\caption{Overview of behaviour of the SPM with pinned spins. 
(a)~Time-dependent correlation function $C(t)$ at inverse temperature $\beta=3$.
(b)~The long-time overlap $q=\lim_{t\to\infty}C(t)$ as a function of pinning
fraction $f$ for four temperatures (the symbols are the same as those in panel e).
(c)~Time-dependent four-point susceptibility $\chi_4(t)$ at $\beta=3$.
(d)~Long-limit limit of $\chi_4$, for various temperatures.
(e)~Behaviour of the relaxation time on varying $f$ at $\beta=3$.
(f)~Behaviour of the maximal relaxation time $\tau_\mathrm{max}=\max_f \tau(f)$,
normalised by the bulk relaxation time $\tau_0=\tau(f=0)$, as a function
of inverse temperature.  In practice $\tau_\mathrm{max}\approx \tau(f^*)$.
The straight line is $\tau_\mathrm{max}/\tau_0 \propto \ee^\beta$ which would
imply $\tau_\mathrm{max}\propto \ee^{4\beta}$.  It is clear from the upwards
curvature of the numerical data that $\tau_\mathrm{max}$
has a super-Arrhenius temperature dependence.}
\label{fig:spm-qt}
\end{figure*}

An overview of the influence of pinned spins on the SPM is shown in 
Fig.~\ref{fig:spm-qt}.
In Fig.~\ref{fig:spm-qt}a we show that as $f$ is increased from zero,
the long-time limit of $C(t)$ increases monotonically, since the frozen 
spins tend to maintain the system close to its initial state,
preventing full relaxation of the autocorrelation function of the mobile spins. 
Thus, we confirm that by randomly freezing
spins, the system crosses over from `bulk' ergodic relaxation 
for $f=0$ to a nearly `frozen' state at large $f$. Our goal
is now to characterize this crossover further.

In Fig.~\ref{fig:spm-qt}b, we show the evolution of 
$q= q(f,T)$ with the fraction of pinned spins
and temperature.
As expected from Fig.~\ref{fig:spm-qt}a, the static value of the 
overlap increases monotonically with $f$ at any given temperature.
More interesting is the
temperature dependence of the $q(f,T)$ curves. We find that $q$
increases rapidly with decreasing temperature at constant $f$. 
This implies that for lower temperatures, a smaller amount of random pinning is required 
to localize the system in a single state.
The interpretation is that the system
has a greater degree of amorphous order at low temperatures.

Looking more closely at the $f$-dependence of 
$q$, the data in Fig.~\ref{fig:spm-qt}b indicate that 
$q(f,T)$ has an inflexion point at a characteristic 
value of the pinning fraction, $f=f^*$, so that the susceptibility 
$(\partial q/\partial f)$ is small both for small $f$ and for large $f$, 
with a well-defined maximum at $f^*(T)$. 
Thus we find that the bulk-to-localized crossover obtained with 
random pinning can be located 
by measuring the derivative of the static overlap.
However, anticipating the discussion
in Sec.~\ref{sec:conc}, we note that $q(f,T)$ is a smooth function 
of $f$ with no sign of the sharp discontinuity that would be observed 
at a first order phase transition~\cite{cammarota2011}. 

We next turn to fluctuations of the overlap, 
which we quantify via the four-point susceptibility $\chi_4(t)$.
In Fig.~\ref{fig:spm-qt}c, we show the time evolution 
of $\chi_4(t)$ for different values of $f$, at constant temperature.
For $f=0$, the susceptibility 
$\chi_4(t)$ has a peak for $t \approx \tau$, as usual in glassy systems~\cite{bookdh}.
However, two features emerge when 
$f$ is increased. First, the time dependence changes 
dramatically: the maximum in $\chi_4(t)$ shifts to longer times until,
for large $f$, the susceptibility $\chi_4(t)$ is monotonically increasing 
and saturates to a plateau
at long times.  This long time limit corresponds to the static susceptibility
$\chi_4 =\chi_4(f,T)$, proportional to the variance in 
the overlap $q(f,T)$. 
This increasing static susceptibility indicates that 
deviations between the final configuration 
and the initial (reference) configuration appear by cooperative processes 
involving many spins.  Just like $(\partial q/\partial f)$,  
the static susceptibility $\chi_4$
goes through a maximum at the characteristic pinning fraction $f^*$. 

Physically, the interpretation of the behaviour of $\chi_4(t)$ is as follows. 
When $f \ll f^*$, the system is in the bulk regime and easily 
escapes from the reference state through a process that 
is not very different from bulk relaxation. In this case, $\chi_4(t)$ is large 
near $t \approx \tau$, but it is small at long times 
since initial and final states are very different.
When $f \gg f^*$, by contrast, there are so many frozen spins 
that the system is very constrained and few spins 
can relax. While the overlap is large, its fluctuations 
are necessarily quite small. 
For $f \approx f^*$, the number of frozen spins is just large enough 
to maintain the system near its initial state, and 
the overlap exhibits stronger fluctuations because the system
`hesitates' between both possibilities (``{\it should I stay or 
should I go}'').  
We discuss the spatial structure of these correlations in Sec.~\ref{sec:lengths} below. 

We show in Fig.~\ref{fig:spm-qt}d the evolution of
the static susceptibility $\chi_4$ with $f$ for different 
temperatures. As anticipated, the susceptibility 
goes through a maximum whose location and amplitude are strong 
functions of the temperature. On going to lower temperature,
the peak of $\chi_4$ remains located near $f=f^*$ so it shifts towards smaller values of $f$; 
the amplitude of the peak increases rapidly, and its width decreases. 
Thus, 
the crossover between bulk and localized behaviours becomes sharper and more pronounced
at low temperatures. 
A relevant conclusion for supercooled liquids is that the data in  Fig.~\ref{fig:spm-qt}d
show an increasing static susceptibility that measures the growth of amorphous order,
but these data are obtained without any
a priori knowledge of the many-body correlations that are responsible for this order. 
Thus, the random pinning procedure is 
a generic way to measure a growing static susceptibility 
in liquids approaching the glass transition, and offers
a thermodynamic alternative to the measurement of relevant length scales
via dynamic heterogeneity. 

Turning to the dynamic behaviour in the presence of random pinning,
we define a relaxation time $\tau=\tau(f,T)$ from the time 
decay of $C(t)$, via $\frac{C(\tau)-q}{1-q}=1/\ee$.
We compute $\tau$ using the data shown in Fig.~\ref{fig:spm-qt}a
and show the results in 
Fig.~\ref{fig:spm-qt}e. We also find that 
$\tau$ has a non-monotonic behaviour, the relaxation 
being slowest near $f^*$. While the maximum is not very pronounced 
in Fig.~\ref{fig:spm-qt}e, we show in Fig.~\ref{fig:spm-qt}f
that the ratio $\tau(f^*,T)/\tau(f=0,T)$ increases when 
temperature is reduced.
This is consistent with the presence of
increasingly cooperative relaxation mechanisms in the presence of pinned 
spins.  While the SPM without pinning 
has `strong' glass scaling, $\tau(f=0)\sim 
\ee^{3/T}$, we find that $\tau(f^*)$ increases
in a super-Arrhenius (fragile) fashion.  This indicates that the 
relaxation mechanism for the system near $f^*$
is different  
from the bulk mechanism at $f=0$, presumably because the frozen spins act to
frustrate relaxation of the mobile ones.

\begin{figure}
\includegraphics[width=7cm]{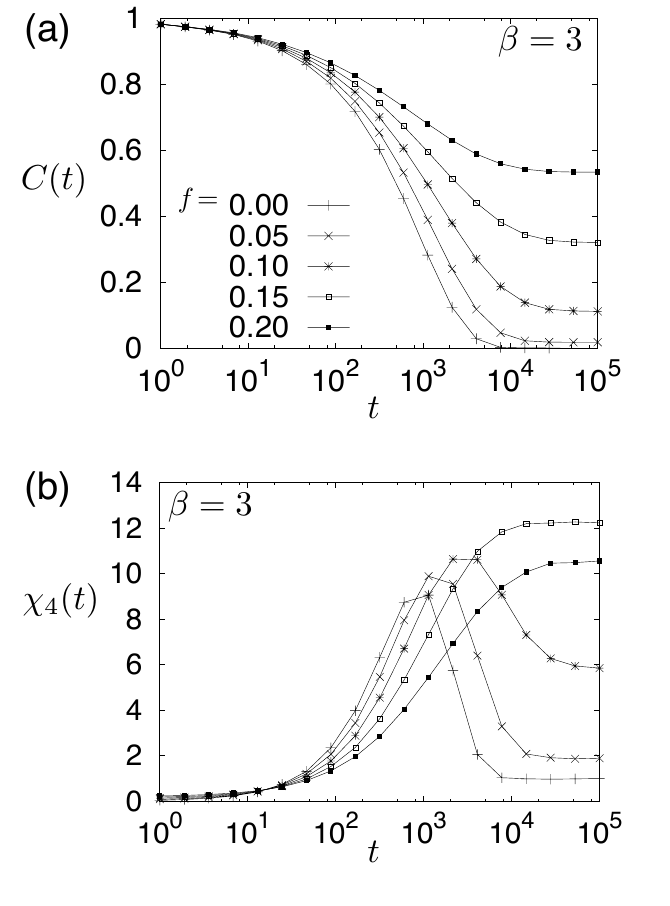}
\caption{Effects of pinning in the TPM at $\beta=3$.  
(a)~Time-dependent correlation function.
(b)~Time-dependent four-point susceptibility.}
\label{fig:tpm-qt}
\end{figure}

We expect many of the results shown for the SPM to be quite generic in 
glassy systems.  Certainly, the behaviour in the TPM is similar.
Figure~\ref{fig:tpm-qt} shows data for $C(t)$ and $\chi_4(t)$ in the TPM
at the representative temperature $\beta=3$. As before, 
$C(t)$ exhibits a plateau at long times which increases with $f$, 
while the static susceptibility $\chi_4$ is 
maximal at some $f=f^*$. Consistent with the SPM,
we also find that $f^*$ decreases at low temperature, that the 
maximum of the static susceptibilities increase, and that the  
time scales at $f^*$ increase. These results
resemble strongly the ones shown in Fig.~\ref{fig:spm-qt}, and are therefore
not shown, for brevity.  However, the $f$-dependence of the relaxation 
time $\tau$
is weaker in the TPM than in the SPM: the ratio $\tau_\mathrm{max}/\tau_0$
does increase systematically on decreasing temperature but it 
takes values in the range
$1-4$ while $\tau_0$ varies over nearly four orders of magnitude.

We emphasize that all of the results presented in
this article are obtained in large systems and we 
have checked that they are free from
finite size effects.  In particular, susceptibilities 
and relaxation times have maxima at 
$f^*$ but the maximum values remain finite even when the thermodynamic
limit is taken, $V \to \infty$. Thus, there are no diverging
correlation times or correlation lengths in this system for any finite 
$f$ or $T$, nor is there any phase transition. 
However, we find that correlation times and susceptibilies
have sharp maxima along a line $f^*(T)$ in the $(f,T)$
phase diagram.

\section{Scaling of lengths}
\label{sec:lengths}

We have shown that varying $f$ in the SPM and TPM reveals 
crossovers at $f^*$, associated
with maxima in susceptibilities and in relaxation times.  
We now discuss how these 
features can be related to correlation lengths in these systems.  
In particular, we 
focus on the scaling of these lengthscales at low temperatures.

\subsection{Visualisation of spatial correlations}
\label{subsec:snapshot}

It is instructive to visualise the spatial fluctuations 
that appear as a result of the random pinning. 
To this end, we consider 
the dynamic propensity~\cite{harrowell2004,harrowell2007}.  (Compared to
visualising the autocorrelations $a_i$ directly, the propensity 
provides continuous functions rather than binary ones, 
and this yields images that better differentiate between regions where relaxation is
frustrated by the frozen spins and those where relaxation can occur.)

To calculate the propensity, 
we take a single representative reference configuration, 
$\bm{s}^\mathrm{A}$, in which a specific set of spins are frozen, and we run 
several long MC trajectories starting from
it.  We calculate the autocorrelation $a_i(t) = s_i(t)s_i(0)$ 
for each unfrozen spin in each trajectory and we average over the 
trajectories to obtain the (site-dependent)
propensities $p_i(t) = \overline{a_i(t)}$.  
For large times these propensities approach limiting
values, $p_i = p_i(t\to\infty)$, 
which depend on the reference configuration $\bm{s}^\mathrm{A}$ 
but not on the time $t$.  The propensities are therefore 
static on-site quantities characterizing the degree 
of freezing of spin $i$ for a given realization
of the random pinning and a given reference configuration.

\begin{figure}
\includegraphics[width=8cm]{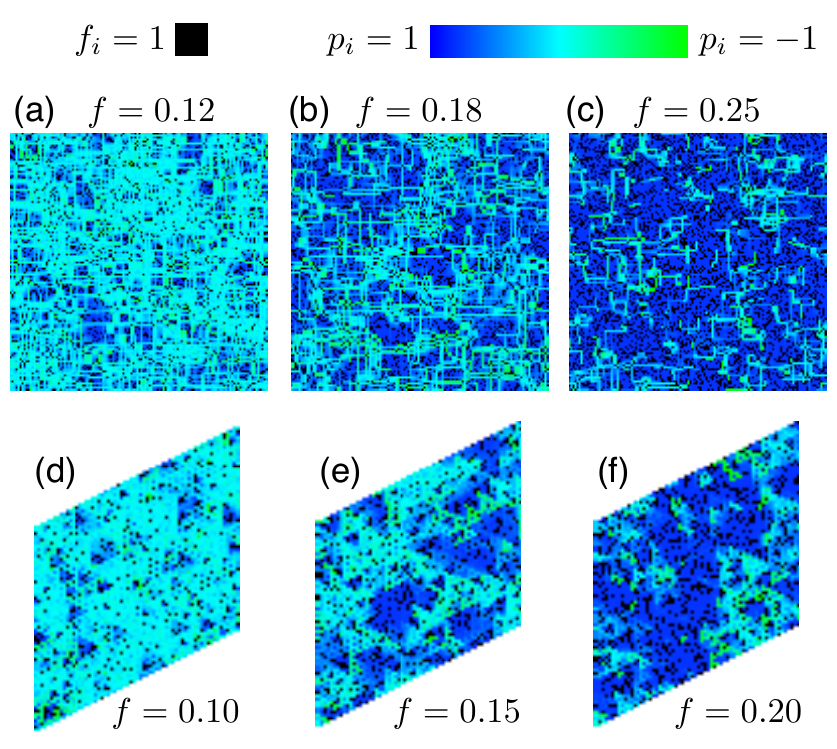}
\caption{Snapshots of the propensities $p_i$ in the long-time limit.
(a-c)~SPM at $\beta=3$, varying $f$.  At this temperature $f^*\approx 0.18$.
(d-f)~TPM at $\beta=3$ for which $f^*\approx0.15$.  
Pinned spins are black while unpinned
spins are color-coded with dark blue for $p_i\approx 1$,
pale blue for $p_i\approx0$ and green for $p_i\approx-1$.
On increasing $f$, the spins become polarised so that $p_i\approx1$ in most cases.
The correlations associated with the propensity are strongest near $f^*$, consistent
with $\chi_4$ being maximal.}
\label{fig:prop}
\end{figure}

Representative results are shown in 
Fig.~\ref{fig:prop}, where pinned spins are shown in black, and 
a blue-green colour coding describes the propensity.
Sites for which the pinned spins cause the configuration to remain near its 
initial state have $p_i\approx 1$ (dark blue) while those
where the pinned spins have little effect have $p_i=0$ (light blue).  
If the frozen spins cause $s_i$ to become polarised in the
opposite direction to $s_i^\mathrm{A}$, one finds $p_i<0$ (green).  

At $f=0$ then $p_i=0$ for all $i$ and the system is 
homogeneous. As $f$ starts to increase (left panel),
then the system acquires regions where the spins 
become polarised and cannot relax any more (coloured 
dark blue in Fig.~\ref{fig:prop}).  This seems to occur in 
small, isolated regions whose size increases with $f$. 
For $f\approx f^*$ (middle panel), these polarized regions percolate 
throughout the system and spatial fluctuations of the propensity 
occur over a large lengthscale. This yields snapshots where  
large regions are strongly polarised while others are unaffected by 
the pinning.  Finally, as $f$ increases further above 
$f^*$ (right panel), most spins are strongly 
pinned, and only few small regions exist where 
motion remains possible. In this regime, the system is strongly 
localized near the reference configuration.

An important observation
is that the lengthscales associated with these 
correlations observed near $f^*$ are much 
longer than the typical distance $f^{-1/2}$ 
between pinned spins.  This is apparent in 
Fig.~\ref{fig:prop}, because the coloured domains 
are clearly much larger than the spacing between (black) pinned 
sites. In other words, each correlated region in these images 
contain very many pinned spins. These observations will be 
quantified below in Sec.~\ref{subsec:f*scal}.

The qualitative description of these images is strongly 
reminiscent of observations made in dynamic heterogeneity 
studies~\cite{bookdh},
except that time has now been replaced by the fraction of pinned 
spins. The images in Fig.~\ref{fig:prop} suggest that a similar behaviour
is found in both square and triangular models but 
that the specific features of the models will be reflected
in the form of the correlation functions.  For example, the SPM is 
characterized by strongly anisotropic correlations,
while correlations appear more isotropic for the TPM, although they do have an underlying fractal structure. 
These observations once again echo previous studies
of the dynamic heterogeneity in these models~\cite{plaq-static2005}

\subsection{SPM: `Bulk' scaling at $f=0$}
\label{subsec:f0scal}

To analyse length scales and their scaling in the SPM,
it is useful to start by considering
static correlations for the `bulk' at $f=0$.
As discussed in Sec.~\ref{sec:models}, the first non-trivial 
correlations involve four spins arranged at the edges of a rectangle
of size $x \times y$.
It is clear from Eq.~(\ref{equ:fourspin}) that lines $xy=\mathrm{const.}$ 
are contours of the static four-point function $\Cs(x,y)$. 
That is, the four-spin correlations are strongly anisotropic, which leads 
to unusual scaling behaviour at low temperature ($c\to0$).  
For example, one may measure correlations at a fixed 
finite angle $\theta$ to the lattice axes (with $\theta \neq0,\pi/2$, etc),
in which case 
\begin{equation}
\Cs(x,y) \simeq G_\mathrm{so}(r\sqrt{c},\theta),
\end{equation}
with $x=r\cos\theta$ and $y=r\sin\theta$, as usual.  
(Explicitly $G_\mathrm{so}(u,\theta)=\ee^{-u^2\sin2\theta}$.)
Since the scaling variable is $u = r\sqrt{c}$, the correlation 
length away from the lattice axes scales as $1/\sqrt{c}$.
(Throughout this section we use the symbol $G$ for scaling functions, 
with the approximate equalities valid
on taking $c\to0$ with the arguments of $G$ held constant.)  

However, 
a larger static correlation length in this system is revealed 
by measuring
$\Cs(x,y)$ along the axes of the square lattice.  For fixed $y$ (of order 
unity) and varying the temperature, one gets
\begin{equation}
\Cs(x,y) \simeq G_\mathrm{sa}(xc,y)
\label{equ:gsxy}
\end{equation}
indicating a correlation length $\xi \sim c^{-1}$, measured along the 
lattice axes.  
(Explicitly $G_\mathrm{sa}(u,y) = \ee^{-2yu}$).

At low temperatures, one may also show that the circular average of 
$\Cs(x,y)$ is dominated by contributions
from near the axis and so it also 
decays on a lengthscale $\xi\sim c^{-1}$, as 
$\Cs(r) \simeq r^{-1} G_\mathrm{sr}(rc)$, which may
be rewritten as 
\begin{equation}\Cs(r) \simeq c\, G_\mathrm{sc}(rc).\end{equation}
That is, the effect of the circular average is to pick up the longest 
of the two lengthscales that appear in $\Cs(r)$.

\subsection{SPM: Real-space scaling at $f=f^*$}
\label{subsec:f*scal}

\begin{figure}
\includegraphics[width=7cm]{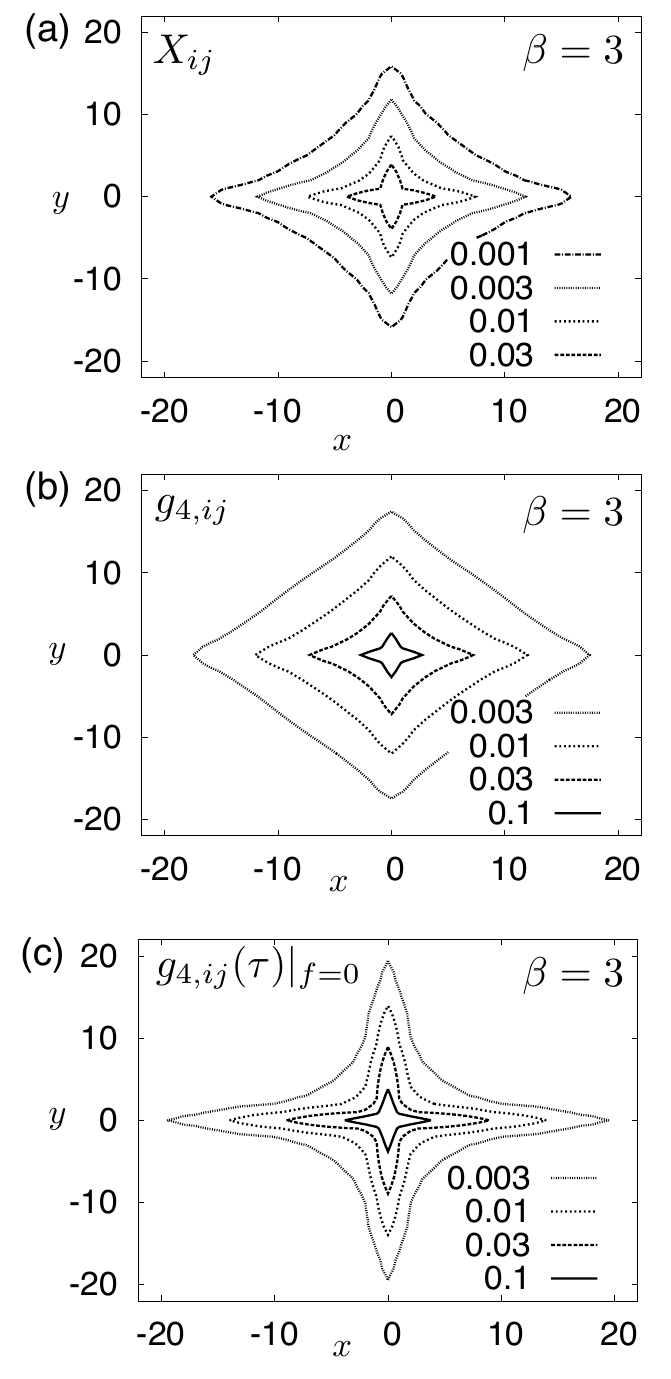}
\caption{Real-space correlations in the SPM.  
(a)~The correlation function $\Xij$ evaluated at $f=f^*$,
in the long-time limit. The co-ordinates $x$ and $y$
give the position of spin $j$ relative to spin $i$.
(b)~The four-point susceptibility $\gij$ at $f=f^*$ and long times.
(c)~The four-point susceptibility at $f=0$, evaluated at the bulk
relaxation time $\tau$.  The scales are the same in all plots, indicating
that all correlations operate over similar length scales.
}
\label{fig:correl-2d}
\end{figure}

In Fig.~\ref{fig:correl-2d}, we show the behaviour of $\gij$ and $\Xij$ 
for $f=f^*$ and a representative temperature $\beta=3$. 
We compare these correlations with
the behaviour of $\gij(t)$ measured at $f=0$ and $t=\tau$, for the same 
temperature.  In all cases, the correlations seem to operate
over a similar lengthscale (the same linear
scale is used for all panels).  The correlation functions
are all strongly anisotropic, 
although we observe slightly different angular dependences 
in each case.  

\begin{figure}
\includegraphics[width=7cm]{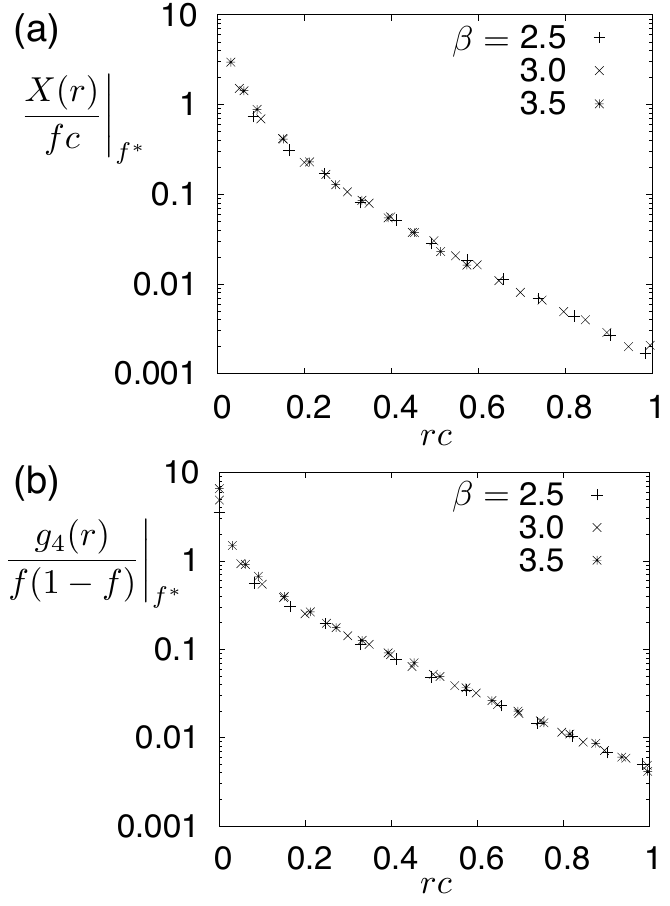}
\caption{Circular averaged correlation functions in the SPM evaluated
at $f=f^*$ and plotted to illustrate their scaling with temperature.  
(a)~Circular average of $\Xij$.  (b)~Circular average of $\gij$.
In both cases the relevant length scales scale as $\xi\sim c^{-1}$.
At the lower temperatures we show  
data points are only for those $r$ and $\beta$ where numerical uncertainties
are smaller than or comparable to symbol sizes.}
\label{fig:circ-scal}
\end{figure}

The dominance of a single lengthscale in this problem
may be seen from Fig.~\ref{fig:circ-scal} where we show circular averages 
of these correlation
functions, plotted as a function of the scaling variable $rc$.  It was 
shown in Ref.~\cite{plaq-static2005} that
dynamical four-point correlations at $f=0$ and $t=\tau$ collapse as a 
function of this scaling variable.
Here we show that the same behaviour holds 
for $\Xij$ and $\gij$ evaluated at $f^*$. 
%

To describe the scaling behaviour of correlation functions near $f^*$ in the SPM,
we make an
ansatz for the circular 
averaged three-point function $X(r)$:
\begin{equation}
X(r)|_{f^*} \approx cf^* G_{X} (rc).
\label{equ:xrscal}
\end{equation}
The choice of scaling variable $rc$ indicates that the dominant correlation 
length in the system scales
as $\xi\sim c^{-1}$. To understand the prefactor $cf^*$, note that
the ratio $\Xij/f$ quantifies
the effect of freezing spin $j$ on the autocorrelation 
function at site $i$. Thus, if each frozen spin 
has an $O(1)$ effect then
one would expect $X(r) \approx f G(r)$, which 
explains the presence of $f^*$ as a prefactor in
Eq.~(\ref{equ:xrscal}). 
The extra prefactor of $c$ in Eq.~(\ref{equ:xrscal}) has two possible
interpretations, which are hard to discriminate on 
the basis of our numerical results.
First, if the correlation function $X(x,y)$ is largest near the 
lattice axis, and if these on-axis correlations are $O(1)$ and
dominate the circular average then one arrives at $X(r)/f \approx c\, 
G_X(r)$, as in the case of the static
function $\Cs(r)$.  However, a second explanation could 
be the presence of off-axis correlations of strength 
$O(c)$, which would lead to the same prefactor $c$ in
Eq.~(\ref{equ:xrscal}).

Our ansatz for the low-temperature scaling of of $g_4$ is
\begin{equation}
\left.g_4(r)\right|_{f^*} \approx f^* (1-f^*) G_4(rc).
\label{equ:grscal}
\end{equation}
Again, the scaling variable $rc$ indicates that the correlation length 
scales as $\xi\sim c^{-1}$.  In Fig.~\ref{fig:circ-scal}, we plot
$g_4/[f^*(1-f^*)]$ as a function of the scaled variable $rc$,
a procedure which nicely collapses our data.  
The prefactor $(1-f^*)$ is irrelevant for the purpose of 
scaling in the low temperature limit.
However, it is natural from a 
physical point of view because $g_{4,ii}\propto(1-f)$, and  
we do find that it improves
the data collapse in the studied range of temperatures.
The physical
interpretation of the prefactor $f^*$ in Eq.~(\ref{equ:grscal}) is 
not immediately clear.
We note that Eqs.~(\ref{equ:xrscal}) and (\ref{equ:grscal}) are 
together consistent with 
Eq.~(\ref{equ:g4conv}), which holds if the correlations of the $a_i$ are 
directly attributable
to individual frozen spins $f_j$.  More support for this can be obtained 
through a direct numerical evaluation of the right hand side of 
Eq.~(\ref{equ:g4conv}) which has the same dependence on $ij$ 
as $\gij$, but is smaller by factor
close to $5$, independently of the temperature. 
As a result, $\chi_4$ and its bound $\Delta \chi_4$ scale in the same way,
but differ by a prefactor.  

To conclude, we have shown robust evidence that $\gij$ and $\Xij$ in the SPM 
are both
controlled by the same lengthscale $\xi\sim(1/c)$.  The scaling of the 
prefactors in these correlations
is less clear, but Eqs.~(\ref{equ:xrscal}) and 
(\ref{equ:grscal}) are consistent with our numerical data. Assuming
that these results do hold, we arrive at the following 
scaling behaviours for the susceptibilities:
\begin{equation}
\frac{\partial q}{\partial f}\bigg|_{f^*} \sim c^{-1}, \qquad \chi_4|_{f^*} \sim 
\Delta\chi_4|_{f^*} \sim c^{-2} f^*. 
\end{equation}

We recall that dynamical correlations $g_4(r,t)|_{f=0}$ at $t=\tau$ are 
controlled by the same
lengthscale $\xi\sim(1/c)$, but that $\chi_4(t=\tau)|_{f=0} \sim c^{-1}$ 
due to the strong 
anisotropy of the correlation function. Combined with the super-Arrhenius 
growth of the relaxation
time shown in Fig.~\ref{fig:spm-qt}f, this difference in the scaling of 
$\chi_4$
emphasizes that the relaxation near $f^*$ is qualitatively different from bulk
relaxation at $f=0$, even if 
the same lengthscale appears in both cases.
In particular, the susceptibility $\chi_4$ at $f^*$ grows more quickly
on cooling than the bulk $\chi_4$, consistent with the observation
that relaxation is slower and more co-operative.  In this respect
increasing $f^*$ is similar to reducing the temperature.  However,
in contrast to decreasing $T$, there is no evidence for an
increasing length scale as $f$ is increased.

%

\subsection{TPM: Real-space scaling at $f=f^*$}
\label{subsec:f*scalTPM}

\begin{figure}
\includegraphics[width=7cm]{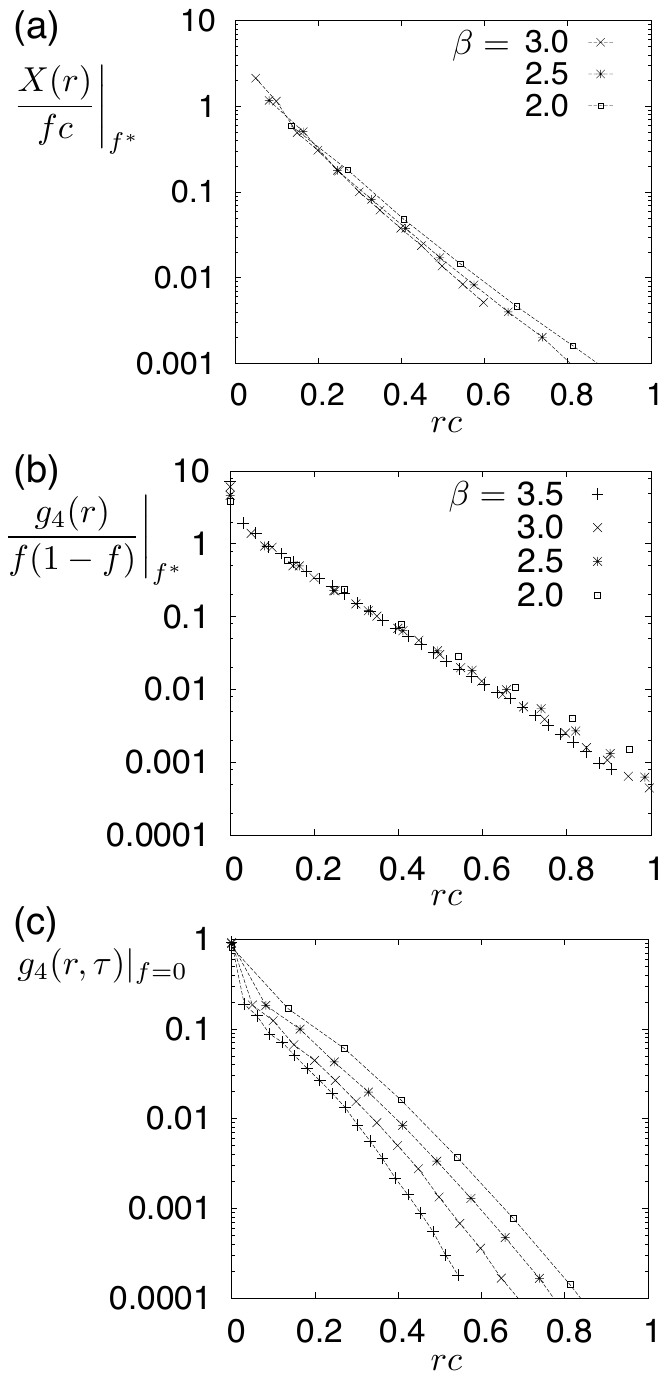}
\caption{
Circular-averaged correlations in the TPM.
(a)~Circular average of $\Xij$ at $f^*$, showing a length scale $\xi\sim c^{-1}$.
(b)~Circular average of $\gij$ at $f^*$, showing a similar length scale $\xi\sim c^{-1}$.
(c)~Circular average of $\gij(t)$ evaluated at $f=0$ and $t=\tau$.  This correlation
function clearly has a different scaling to those evaluated at $f^*$: it was shown in~\cite{plaq-caging2005}
that $\xi\sim c^{-1/\df}$ in this case.
}
\label{fig:tpm-grscal}
\end{figure}

For the triangular plaquette model, the isotropic images shown 
in Fig.~\ref{fig:prop} lead us to  compute
circularly averaged correlation functions directly. The results are presented
in Fig.~\ref{fig:tpm-grscal}, showing scaling with temperature. 
We find that our data are most 
consistent with
\begin{equation}
X(r)|_{f^*} \approx cf^* G_{X\mathrm{T}}(rc),
\label{equ:tpm-xrscal}
\end{equation}
and
\begin{equation}
g_4(r)|_{f^*} \approx f^*(1-f^*) G_{4\mathrm{T}}(rc).
\label{equ:tpm-grscal}
\end{equation}


The most striking feature of Eqs.~(\ref{equ:tpm-xrscal}) 
and (\ref{equ:tpm-grscal}) is that the scaling variable 
$rc$ is the same as that found in the SPM. This suggests
that correlations at $f^*$ extend over a length which 
scales as $\xi \sim 1/c$.
This is surprising because the point-to-set length for the 
TPM at $f=0$ does not scale as $c^{-1}$, nor does the 
dynamic correlation length. 
For dynamical correlations at $f=0$ and $t=\tau$ then it is 
known~\cite{plaq-caging2005} 
that $g_4(r,\tau)|_{f=0}\approx G_{4\mathrm{T}}(rc^{1/\df})$, 
such that both static and dynamic lengths scale as
$\xi(f=0) \sim  c^{-1/\df} \sim c^{-0.631}$.
In Fig.~\ref{fig:tpm-grscal}c, we show that these two scaling forms can be 
clearly differentiated over the temperature range shown,
since $g_4(r)|_{f=0}$ does not collapse as a function of $rc$, 
as expected. It is 
therefore clear that a new lengthscale $\xi \sim c^{-1}$ 
appears in the TPM near $f=f^*$, which is 
longer than any static or dynamic correlation 
length found previously for the bulk at $f=0$. 

As in the SPM, the physical interpretation of the scaling prefactors in 
Eqs.~(\ref{equ:tpm-xrscal}) and (\ref{equ:tpm-grscal}) is not
clear.  The scaling laws we have proposed indicate that the bound 
$\Delta\chi_4$ scales in the same
way as $\chi_4$, although $\Delta\chi_4$ is significantly smaller than 
$\chi_4$ in the TPM.
We also note that while the scaling forms in Eqs.~(\ref{equ:tpm-xrscal}) and 
(\ref{equ:tpm-grscal}) 
in the TPM are the same as in Eqs.~(\ref{equ:xrscal}) and (\ref{equ:grscal}) 
in the SPM, we do not see any a priori reason for this result.
In particular, the spatial structure 
of the correlations are quite different in both cases.

While the appearance of a new lengthscale near $f^*$ makes the TPM 
different from the SPM, we emphasize that the relaxation
mechanism changes qualitatively near $f^*$ in both models.  In the SPM, 
this appears as a larger relaxation time
and a larger susceptibility without any increase in the length; in the 
TPM the lengthscale, time scale and susceptibilities are
all different from their values at $f=0$. 

\section{Square plaquette model:  Weak and strong pinning}
\label{sec:weak-strong}

We have discussed the scaling of length and time scales at $f^*$, 
as temperature is reduced.  In this
section, we consider how $f^*$ depends on temperature.  We focus 
on the SPM for which analytic calculations provide useful insight. 

\subsection{Small-$f$ limit}

We concentrate on the behaviour of the correlation function $\Xij$.  
We define a parameter $\mu$ by $f=(1+\ee^{\mu})^{-1}$
so that the limit of small $f$ is equivalent to a limit of small $\ee^{-\mu}$.
The correlation function $\Xij$ 
has a series expansion in powers of $\ee^{-\mu}$ given by
\begin{multline}
X_{ij} = \frac{1}{Z_\mathrm{f}} \left[ \langle a_i f_j \rangle_0 + 
\ee^{-\mu} \sum_k \langle a_i f_j \rangle_k \right.
\\
 + \left. \ee^{-2\mu} \sum_{k<l} \langle a_i f_j \rangle_{kl} + 
O(\ee^{-3\mu}) \right] - qf(1-f), 
\label{equ:Xseries}
\end{multline}
where $Z_\mathrm{f}=(1+\ee^{-\mu})^V$ and $\langle \cdots \rangle_{kl\dots}$ 
is an average
in a system where spins $(k,l,\dots)$ are pinned and all other spins are
 mobile (unpinned).  
For details, see Appendix~\ref{app:pin}, particularly 
Eq.~(\ref{equ:fexp-general}).
The factor of $q$ in Eq.~(\ref{equ:Xseries})
must be obtained by a separate series expansion over $f$.

\begin{figure}
\includegraphics[width=8cm]{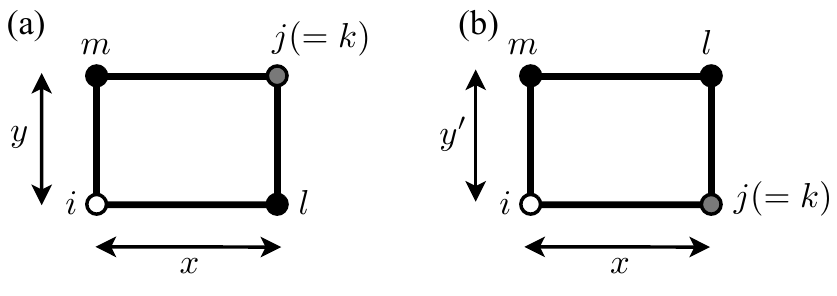}
\caption{Two diagrams representing contributions to $\Xij$ in the SPM, at 
order $f^3$.  The white circle
represents spin $i$, the grey circle spin $j$, and the positions of spins 
$klm$ are to be summed,
subject to the constraint that $j$ is equal to one of $klm$.  
(a)~Leading contribution to $X(x,y)$, assuming $x,y>0$.
(b)~Leading contribution to $X(x,0)$: summing over the positions of spins 
$l$ and $m$ corresponds
to a sum over the dimension $y'$.
}
\label{fig:spm-diag}
\end{figure}

Assuming $i\neq j$, symmetries of the SPM imply that the 
first non-zero term in the expansion is at third order,  
\begin{equation}
\ee^{-3\mu} \sum_{k<l<m} \langle a_i f_j \rangle_{klm}.
\end{equation}  
The factor $f_j$ means that the average is zero 
unless $j$ is equal to one of $k$, $l$, or $m$.    
As shown in 
Fig.~\ref{fig:spm-diag}, the correlations may be calculated in a 
diagrammatic expansion.  Spin $i$ is shown
as a white circle and has a fixed position.  Spin $j$ is shown as a 
grey circle: its position is fixed, and we also
have the constraint that one of the frozen spins $k$, $l$, or $m$ 
coincides with $j$. 
Equation (\ref{equ:Xseries}) shows
that we must sum over the positions of the remaining frozen spins: 
these spins are shown as black circles.
To evaluate the contribution of each diagram to $\Xij$, we use 
Eq.~(\ref{equ:ave-fixed-f}) which shows how
to evaluate expectation values in the presence of a fixed set of frozen spins.
For the SPM, $\langle a_i f_j \rangle_{klm}$ is non-zero only if spins 
$iklm$ lie on the four vertices of a rectangle.  
If the rectangle is of size $x\times y$ then Eq.~(\ref{equ:ave-fixed-f}) 
yields $\langle a_i f_j \rangle_{klm}=\tanh(\beta/2)^{2xy}$.
(The denominator
in (\ref{equ:ave-fixed-f}) has a trivial value $2^{-3}$ in this case, 
because all configurations of the frozen spins are equally likely
and they all have equivalent effects on spin $i$.) 

The leading order behaviour of $\Xij$ stems from 
two distinct cases, as shown in 
Fig.~\ref{fig:spm-diag}.
If spins $i$ and $j$ are in the same row of the square
lattice, with spacing $x$, then we fix $k=j$ and we sum over all sites 
$l$ and $m$ such that the sites $iklm$ form a rectangle.  For a 
rectangle of size $x\times y'$ then Eq.~(\ref{equ:ave-fixed-f}) yields 
$\langle a_i f_j \rangle_{klm}=\tanh(\beta/2)^{2xy'}$.   
Summing over the positions of spins $l$ and $m$, one obtains a geometric
series and the result is, for $x\neq0$:
\begin{equation}
X(x,0) = f^3 \frac{\tanh(\beta/2)^{2|x|}}{1-\tanh(\beta/2)^{2|x|}} + O(f^4)
\end{equation}
where we have defined $X(x,y)=\Xij$, evaluated for spins $i$ and $j$ that are separated
by a vector $(x,y)$.
A similar analysis applies if spins $i$ and $j$ are in the same column of 
the lattice, yielding $X(0,y)=X(y,0)$.

However, if spins $i$ and $j$ are in different rows and columns then we 
fix $l=j$ and we sum over sites $k$ and $m$ such
that sites $iklm$ still form a rectangle.  
There is only one choice for $k$ and $m$ 
in this case, as shown in Fig.~\ref{fig:spm-diag}b.  
If the vector from site $i$ to site $j$
is $(x,y)$ then the resulting rectangle is $x\times y$ in size and,
again, $\langle a_i f_j \rangle_{klm}=\tanh(\beta/2)^{2xy}$ so that,
for $x,y\neq0$,
\begin{equation}
X(x,y) = f^3 \tanh(\beta/2)^{2|xy|} + O(f^4).
\end{equation}

\begin{figure}
\includegraphics[width=7cm]{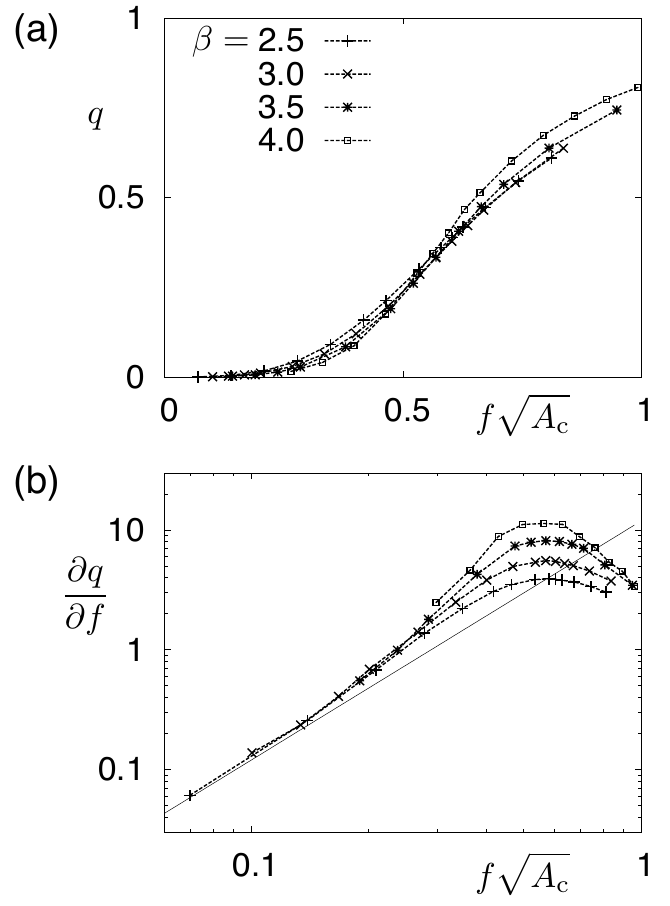}
\caption{Plots of $q$ and $\partial q/\partial f$ for the SPM, 
as a function of the scaling variable $f\sqrt{A_c}$.
(a)~Plot of the overlap $q$ indicating that $f^* \sim A_c^{-1/2}$.
(b)~Slight change in data. The susceptibility $\partial q/\partial f$ is maximal
near $f^*$.  The straight line is the theoretical result (\ref{equ:dqdf})
for the small-$f$ limit.  This approximation applies only when $f\sqrt{A_c}\ll 1$ and does
not capture the peak in $\partial q/\partial f$ near $f^*$.
}
\label{fig:fscal}
\end{figure}

Collecting all these results and summing over the volume,
Eq.~(\ref{equ:xchi}), one finally obtains the 
leading order behaviour of the three-point susceptibility,  
namely 
\begin{equation}
\frac{\partial q}{\partial f} = 12 A_c(T) f^2 + O(f^3),
\label{equ:dqdf}
\end{equation}
where 
\begin{equation}
A_c(T) = \sum_{x=1}^\infty \frac{\tanh(\beta/2)^{2x}}{1-\tanh(\beta/2)^{2x}} 
\simeq \frac{\ee^\beta}{4} [ \beta + O(1) ].
\end{equation} 
The final approximate equality holds for small $c$ (i.e., at low 
temperature) and follows because 
$\sum_k (1-\delta)^k/(1-(1-\delta)^k) \approx (1/\delta) 
[ \log(1/\delta) + O(1) ]$ as $\delta\to0$.
In Fig.~\ref{fig:fscal}a we show that the result in Eq.~(\ref{equ:dqdf}) 
holds very well for small values of 
$f \sqrt{A_c}$, but breaks down for larger $f$,
where higher order terms in the expansion 
also contribute, as discussed below.

The physical interpretation of this small-$f$ result is that adding one extra
frozen spin $j$ affects the autocorrelation $a_i$ over an area of linear 
size $\xi \sim c^{-1}$, but the anisotropy of correlations 
imply that the correlation volume is
$v \sim A_c ~\sim (-\ln c) / c$, 
much smaller than the naive assumption $v \sim \xi^{2} \sim 1/c^2$.
The strength of the response on adding the frozen spin is small (proportional to $f^2$) 
but the lengthscale controlling $\Xij$ 
is large and independent of $f$ as $f\to0$, and the  
sum in Eq.~(\ref{equ:xchi}) is dominated by correlations 
close to the lattice axes. 

Using Eq.~(\ref{equ:dqdf}), it is then easy to 
integrate $\partial q/\partial f$ to obtain the low-$f$
behaviour of the overlap, 
\begin{equation}
q = 4 A_c f^3 + O(f^4).
\end{equation}
It is interesting to remark that 
$A_c = \sum_{xy(\geq 0)} \Cs(x,y)^2$, which highlights the
fact that in the limit where the randomly frozen spins are dilute, 
the static overlap and susceptibilities simply capture the 
most trivial behaviour of the bulk system.
In a supercooled liquid where two-body correlations do not vanish, 
one would expect the overlap to be proportional to $f$ 
for small-$f$, with a prefactor directly given by the pair correlation
function, see Eqs.~(\ref{equ:Xseries}) and (\ref{equ:qflin}). It is only by going beyond
the leading order in $f$ one reveals the relevant higher-order correlations 
responsible for amorphous order~\cite{berthier-kob-pts,tarjus2011pin}.

\subsection{SPM: Behaviour near $f=f^*$}
\label{subsec:fstar-scal}

\begin{figure}
\includegraphics[width=8cm]{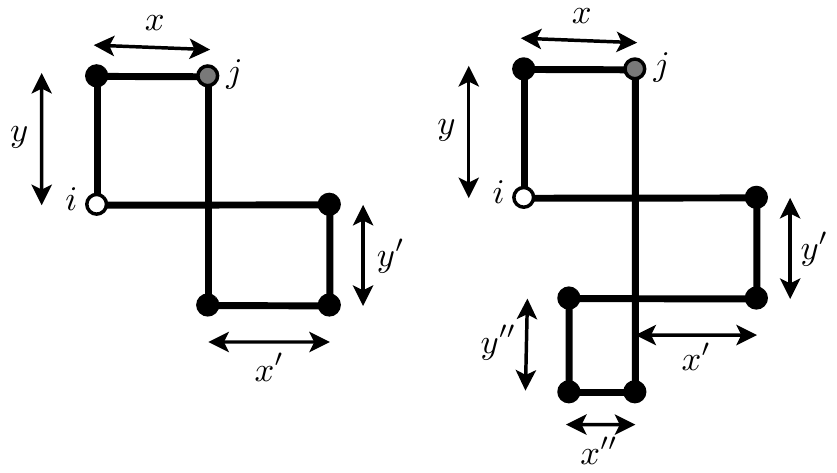}
\caption{Two representative diagrams that contribute to $\Xij$ in the SPM, 
at order $f^5$ (left) and $f^7$ (right). 
The dimensions $x$ and $y$ are fixed by the positions of spins $i$ and $j$ 
while the primed dimensions
are to be summed over.  These sums yield factors of $A_c$ (left) and $A_c^2$ 
(right) in the contributions of
these diagrams to $\Xij$.
}
\label{fig:spm-diag2}
\end{figure}

To analyze the behaviour near $f=f^*$, we now 
consider higher-order terms in 
the expansion over $f$.
In Fig.~\ref{fig:spm-diag2}, we show two further contributions to $X_{ij}$, 
in the case where $i$ and $j$
are not in the same row or column.  
These contributions appear at order $\ee^{-5\mu}$ and $\ee^{-7\mu}$ 
in Eq.~(\ref{equ:Xseries}), respectively.  For these
diagrams, Eq.~(\ref{equ:ave-fixed-f}) yields $\langle a_i f_j\rangle_{kl\dots} 
= \tanh(\beta/2)^{2\cal A}$ where $\cal A$ is the total area of
the rectangular regions enclosed by the solid lines in the diagrams.  
(The denominator in Eq.~(\ref{equ:ave-fixed-f}) is again
trivial for these diagrams, due to symmetries of the SPM.)  Summing over 
the positions of the frozen
spins yields a factor of $A_c$ for each rectangle, except for the 
rectangle whose location is fixed
by the positions of $i$ and $j$. 
These two diagrams therefore
contribute to $\Xij$ as $f^3 (f^2A_c) \tanh(\beta/2)^{2xy}$ and 
$f^3(f^2A_c)^2 \tanh(\beta/2)^{2xy}$, respectively.
Of course, there are various other contributions at these orders that scale 
in the same way. Constructing
higher order diagrams similar to those in Fig.~\ref{fig:spm-diag2},
one may identify a series of positive terms 
which are all proportional to powers of 
$f^2 A_\mathrm{c}$. For small $c$, we
expect these to be the largest terms at each order.  
These considerations   
clearly motivate the use of $f \sqrt{A_c}$ as a 
scaling variable for this expansion.
Of course, if $f^2 A_c$ is not a small number, then the leading
terms in the small-$f$ expansion do not give a good approximation to 
the correlation function.

In Fig.~\ref{fig:fscal} we plot $q$  and $\partial q/\partial f$
as a function of the scaling variable $f \sqrt{A_c(T)}$. 
We find that the crossover $f^*$,
which corresponds to the maximum observed for 
$\partial q/\partial f$,  
does indeed scale as $A_c^{-1/2}$, so that finally
\begin{equation}
f^*(T) \approx \sqrt{\frac{-\ln c}{c}} \sim \sqrt{T} \exp \left(-
\frac{1}{2T} \right).
\end{equation}  
This result shows that the `localization' crossover at $f^*$ 
occurs along a line $f^*(T)$ in the $(f,T)$ phase diagram,
with $f^*\to0$ as $T\to0$.

Also, the numerical data indicate  that results 
from the small-$f$ limit such as Eq.~(\ref{equ:dqdf})
are applicable only when $f\sqrt{A_c}\ll 1$ and break down for $f\simeq f^*$.
As suggested above, 
this strongly suggests 
that the maxima in $\partial q/\partial f$ and $\chi_4$ 
have their origin in non-trivial 
``many-body'' effects that are not captured by the low-order expansion 
about the dilute limit.  

Physically, the interpretation is that the relaxation mechanism near 
to $f^*$ is qualitatively
different from the bulk relaxation at $f=0$.  Perturbation theory in $f$ 
is not sufficient
to capture this new mechanism: a non-perturbative approach would be 
necessary to make
further analytic progress, presumably by summing infinite subsets of 
diagrams in the expansion
of correlation functions. In this model, $f^*\to 0$ when $T\to0$, indicating 
that the regime where perturbation
theory is valid becomes vanishingly small near the glass 
transition of the model (which takes place at $T=0$). 

\section{Discussion}
\label{sec:conc}

We have studied the effect of random pinning 
in the context of two finite dimensional 
spin models with plaquette interactions. By increasing the 
fraction $f$ of pinned spins at a fixed low temperature,
we have discovered the existence of a temperature-dependent crossover 
between
bulk-like relaxation at small $f < f^*(T)$, and a nearly 
localized amorphous state at large $f>f^*(T)$. The study
of static correlation functions and susceptibilities 
in the presence of random
pinning directly reveals the existence of growing
amorphous order on cooling. This growth appears 
through the large lengthscales that
can be measured by $\Xij$ and $\gij$, and by the decreasing values 
$f^*(T)$ required to keep
the system localized in a single state at lower temperature. 
Moreover, these measurements 
do not require a priori knowledge of the specific type of order
that sets in at low temperature. These results therefore 
demonstrate that the main objective underlying the measurement
of point-to-set correlation functions is fulfilled in plaquette models.   

The lengthscales that we measure show scaling
behaviour at low temperatures. We have emphasized that while low-order 
terms in the expansion
over $f$ are related to static correlation functions of the bulk 
system at $f=0$, the length
and time scales that we observe near $f=f^*$ are related to 
nonperturbative effects, and analytic calculations
of lengthscales in that regime would seem to require a new 
approach beyond those given here.

Similarly, we have shown that 
length and time scales near $f^*$ in the plaquette models are 
not related in the same way as they are at $f=0$. 
Increasing $f$ in the SPM, we found an increasing time scale and a 
growing susceptibility, but without any increasing
length scale. In the TPM, the length scale increases as $f$
is increased but the change in time scale is very mild, in contrast 
to the strong dependence observed at $f=0$ for this model.

We have also emphasized that although static and dynamic 
correlations are strongly enhanced near $f^*$ as compared to the bulk, 
length and time scales are finite at $f^*$, after taking
the thermodynamic limit at any non-zero temperature.  This  
implies that random pinning does not induce any kind of phase 
transition in plaquette models.  
It is perhaps unsurprising that these models do not 
exhibit an ideal glass transition in the $(f,T)$ 
phase diagram, since no transition occurs 
at finite temperature in the bulk at $f=0$ either. 
However, the sharp crossovers we have revealed in plaquette models 
represent non-trivial new results, because they have no 
counterparts in the bulk systems at $f=0$.

Given that neither mode-coupling theory nor RFOT theory
represent accurate descriptions of plaquette models 
in the bulk, we do not expect these approaches to 
account for the effect of random pinning either. 
Thus, we argue that the results 
obtained within plaquette models 
provide a useful alternative reference point 
for interpreting simulation data
for more realistic models of supercooled liquids.
In particular,  the absence of any phase transition
at $f^*$ indicates that such transitions may not be generic in
glassy systems with pinning. 

In this respect, it is instructive to compare the results we find here 
with the predictions of RFOT theory~\cite{cammarota2011}. 
A central quantity
in this theory is the configurational entropy, which measures
the diversity of long-lived metastable  states.
Assuming that RFOT applies in supercooled liquids,
results for model systems and renormalisation group 
calculations~\cite{cammarota2011} 
indicate that the configurational entropy density $s_c$
is well represented by 
\begin{equation}
s_c(T,f) \approx s_c(T,0) - f Y(T).
\label{equ:scf}
\end{equation}
In 3 dimensions and above, this leads to a phase transition 
at $f^*(T)\approx Y(T)/s_c(T,0)$.
In addition, RFOT predicts that a lengthscale $\xi_\mathrm{PTS}$ grows 
as $s_c(f,T)$ approaches zero, 
and that the relaxation mechanism at finite $f$ involves cooperative 
rearrangements over the lengthscale $\xi_\mathrm{PTS}$,
much as in the bulk.
We emphasise that $s_c$ is related to metastable states and may not be 
obtained 
from the statistics of minima on the system's energy landscape.
A precise definition of $s_c$ in finite-dimensional 
systems is slightly problematic since
all metastable states have finite lifetime in that case.  However, one 
may follow the procedure of Ref.~\cite{kurchan2011},
as long as the timescale associated with structural relaxation is 
well-separated from all microscopic timescales.

Turning to the plaquette models, the geometric construction of
Ref.~\cite{kurchan2011} indicates that
$s_c$ decreases as $f$ increases, just as in RFOT theory.
However, it is clear from Fig.~\ref{fig:prop} that even for $f>f^*$ 
there are sets of spins that may rearrange cooperatively,
which ensures that $s_c$ does not vanish at $f^*$.
Thus, while $s_c$ presumably decreases sharply near $f^*$, 
it does not drop to zero as predicted by 
Eq.~(\ref{equ:scf}). For supercooled liquids, RFOT predicts 
instead localization in a single state for $f>f^*$, 
so that if the analysis of Sec.~\ref{sec:lengths} were repeated
for those systems then regions where cooperative motion is possible 
should be forbidden for $f>f^*$, and
the light-coloured regions shown in Figs.~\ref{fig:prop} would be 
completely absent.  In plaquette models, the existence of such regions
restores a finite configurational entropy density above $f^*$ and the 
proposed phase transition is avoided.  
It is unclear whether such strong spatial fluctuations can be present
in supercooled liquids, and whether they are properly captured 
by renormalization group treatments~\cite{cammarota2011}.
This remains an area of ongoing research~\cite{cammarota2011-rg,castellana2011,moore2011-cm,angelini-cm}.

Thus, the plaquette models illustrate that even if the detailed 
predictions of RFOT do not apply, systems where configurational
entropy decreases on pinning can be generically expected to exhibit 
increased cooperativity 
on increasing $f$, accompanied by growing timescales 
as well as growing lengths and/or growing static susceptibilities.  
One may also expect crossovers that sharpen and move to small $f$ on cooling. 
To this extent, the plaquette models are broadly consistent with 
published numerical results for particle 
models~\cite{kim2003-pin,kim2011-pin,procaccia2011-pin,berthier-kob-pts,tarjus2011pin}, although the $f$-dependence of the relaxation time appears much  
weaker in the plaquette models in comparison with model 
liquids.  

On the other hand, the most striking prediction of
RFOT theory is the presence of a phase transition at finite $f^*$. 
This transition remains to be found
numerically, as the behaviour for large values of $f$ and low temperatures 
has not been investigated in much detail so far. 
In the absence of such a transition, we argue that 
the detailed RFOT scaling predictions for length
and time scales must be tested directly in order to 
substantiate the theory and distinguish 
it from a more general picture of
increasing cooperativity in the presence of pinning. Testing these 
predictions remains however a very challenging task, especially in 
the absence of direct measurements of the configurational entropy. 
We therefore conclude that while pinning particles is an 
interesting new method
of measuring amorphous order and its growth upon cooling, 
it does not necessarily resolve the central problem of how to 
test the fundamental assumptions of RFOT theory by practical 
measurements. Nevertheless, we hope future studies 
will investigate further the effect of random pinning  
in supercooled liquids, especially in the relatively unexplored  
regime of strong pinning.  

\begin{acknowledgments}

We thank G. Biroli, D. Coslovich, W. Kob, and G. Tarjus 
for useful discussions. R. L. J. thanks the
EPSRC for financial support through grant EP/I003797/1.

\end{acknowledgments}

\begin{appendix}

\section{Correlations in systems with random pinning}
\label{app:pin}

In this appendix, we discuss some general results for spin systems 
in the  presence of pinning.

\subsection{Ensemble dependence of $\chi_4(t)$}

In the systems considered here,
the set of frozen spins $\bm{f}$ remains constant as the dynamics 
proceeds. This results in ensemble-dependent 
susceptibilities~\cite{berthier2005-science,berthier2007-jcp1,berthier2007-jcp2}, such as
\begin{equation}
\chi_4(t) = \chi_{4F}(t) + \Delta \chi_4(t),
\label{equ:chi4ens}
\end{equation}
where $\chi_{4F}(t)=[(1-f)V]^{-1}\langle \delta \QQ(t)^2 \rangle_f$ 
is evaluated in a `restricted' ensemble
with a fixed number of frozen spins.  Analysis of such 
ensemble-dependence can be useful for understanding
how the time-independent variables (the frozen spins in this case) 
influence the time-dependent ones.

The difference term $\Delta\chi_4(t)$ may be derived as 
in Ref.~\cite{berthier2005-science} or equivalently
following Ref.~\cite{berthier-jack2007}. We 
write $\QQ-\langle \hat{Q}\rangle = (\QQ-\langle \QQ \rangle_f) 
+ ( \langle \QQ \rangle_f -\langle \QQ \rangle)$ where $\langle \cdots 
\rangle_f$ is an average with fixed $\Nf$, as above.
Substituing into Eq.~(\ref{equ:chi4}), we note that if the restricted 
ensemble has $\Nf \approx fV$ frozen spins then $\langle \QQ 
\rangle_f -\langle \QQ \rangle 
\approx V^{-1} (\partial \langle \QQ \rangle/\partial f) (\Nf - fV)$ 
which gives $\langle \delta \QQ(t)^2 \rangle = \langle \delta 
\QQ(t)^2 \rangle_f
 + V^{-2}(\partial \langle \QQ \rangle/\partial f)^2 \langle 
(\delta\Nf)^2 \rangle$ (the equality is exact
in the limit of large system size $V$, since the fluctuations of 
$\Nf$ are small in that case).  Noting that $\langle (\delta\Nf)^2 
\rangle = Vf(1-f)$ and
 $\partial \langle \QQ \rangle/\partial f
=V[(1-f)\partial C(t)/\partial f - C(t)]$ then the result 
(\ref{equ:deltachi}) follows.

Since $\chi_{4F}(t)$ and $\Delta \chi_4(t)$ are both
non-negative then $\chi_4(t)\geq \Delta \chi_4(t)$.
If this bound is saturated, this means that the
correlations between the $a_i$ are directly attributable to the 
influence of individual $f_j$.  In particular, a sufficient
condition for $\chi_4(t) = \Delta \chi_4(t)$ is that for a fixed 
choice $\bm{f}$ of frozen spins,
the autocorrelations $a_i$ are all independent and respond linearly 
to the $f_j$, so that
$\langle a_i(t) \rangle_{\bm{f}} = \langle a_i(t) \rangle_0 + \sum_{k} 
f_k U_{ik}(t)$ where $U_{ik}(t) = X_{ik}(t)/[f(1-f)]$ is  assumed
independent of $f$.
Physically, this condition means 
that the site-to-site fluctuations of the $a_i$ depend only on the 
frozen spins, and the effect of each frozen spin is independent.

In addition,
independence of the $\langle a_i \rangle_{\bm f}$ means that 
$\langle a_i a_j \rangle_{\bm f}=\langle a_i\rangle_{\bm f}\langle a_j 
\rangle_{\bm f}$.
Following Ref.~\cite{krak2010} we use an overbar to indicate the average 
over $\bm{f}$, so the definition of the four-point function is
$\gij = \overline{\langle a_i a_j \rangle_{\bm f}} - \overline{\langle 
a_i \rangle_{\bm f}}^2$.  Hence,
$\gij = \sum_{kk'} [\overline{f_k f_{k'}} - f^2 ] U_{ik} U_{jk'}$, and 
since the frozen spins are chosen independently
one has $\overline{f_k f_{k'}}-f^2=f(1-f)\delta_{kk'}$.  The resulting 
expression for $\gij$ is given in Eq.~(\ref{equ:g4conv}) of the main
text, where  $\gij$ appears as a convolution of $\Xij$ with
itself. 

We note that for Ising spin variables $s_i$, it is not possible to 
satisfy $\langle a_i a_j \rangle_{\bm f}=\langle a_i\rangle\langle a_j 
\rangle_{\bm f}$
in the case $i=j$. However, assuming that $\chi_4$ is dominated by 
collective behaviour and not single-site fluctuations, one 
still expects Eq.~(\ref{equ:g4conv}) to hold as an approximate equality 
if the $a_i$ are primarily determined by a linear response to the $f_i$.

\subsection{Long time limit}

As discussed in Sec.~\ref{sec:pinning}, 
long-time limits of correlation functions in systems with pinned spins are
static (thermodynamic) quantities and can be calculated within equilibrium 
statistical
mechanics.  The analysis is similar to that of Krakoviack~\cite{krak2010} 
for particle systems.

We write spin configurations as $\bm{s}=(s_1,\dots, s_V)$ and choices of
frozen spins as $\bm{f}=(f_1,\dots, f_V)$.
The distribution of the initial (reference) configuration $\bm{s}^\mathrm{A}$
is
$ 
P_1(\bm{s}^\mathrm{A}) = \ee^{-\beta E(\bm{s}^\mathrm{A})}/Z_1
$
where $E(\bm{s}^\mathrm{A})$ is the energy of configuration 
$\bm{s}^\mathrm{A}$ and 
the partition function $Z_1$ enforces normalisation.  
The $f_i$ are all independent with $\langle f_i \rangle = f$ and 
so their distribution is
$ 
P_\mathrm{f}(\bm{f}) = \ee^{-\mu \sum_i f_i} /\Zf
$
where $\mu$ is defined through $f= (1+\ee^{\mu})^{-1}$ and $Z_\mathrm{f}$ 
is a normalisation constant.
We denote the number of frozen spins by $\Nf=\sum_i f_i$, noting that
$\Nf+\Nm=V$. 

In the long-time limit, and for fixed $\bm{s}^\mathrm{A}$ and $\bm{f}$,
the final configuration $\bm{s}^\mathrm{B}$ has probability distribution
\begin{equation}
P_2(\bm{s}^\mathrm{B}|\bm{f},\bm{s}^\mathrm{A}) 
 = \frac{1}{Z_\mathrm{B}(\bm{f},\bm{s}^\mathrm{A})} 
\ee^{-\beta E(\bm{s}^\mathrm{B})} \prod_i [ (1-f_i) + f_i 
\delta_{s^\mathrm{A}_i,s^\mathrm{B}_i} ],
\end{equation}
where $Z_\mathrm{B}(\bm{f},\bm{s}^\mathrm{A})$ is a normalisation factor,
defined so that $\sum_{\bm{s}^\mathrm{B}}
P_2(\bm{s}^\mathrm{B}|\bm{f},\bm{s}^\mathrm{A})=1$.
Thus, in the long-time limit, averages $\langle \cdots \rangle$ 
are taken with respect to the distribution
\begin{equation} 
P(\bm{f},\bm{s}^\mathrm{A},\bm{s}^\mathrm{B}) = P_\mathrm{f}(\bm{f}) 
P_1(\bm{s}^\mathrm{A}) P_2(\bm{s}^\mathrm{B}|\bm{f},\bm{s}^\mathrm{A}).
\label{equ:Pall}
\end{equation}  
It may be shown that this distribution is invariant under 
$\bm{s}^\mathrm{A}\leftrightarrow\bm{s}^\mathrm{B}$.  In particular,
this means that the marginal distribution of $\bm{s}^\mathrm{B}$ 
is equal to the Boltzmann distribution $P_1(\bm{s}^\mathrm{B})$.
Physically, this means that structural averages and correlation 
functions are unaffected by the pinning.
Long-time correlations between the $f_i$ and
the $a_i$ may be calculated as averages with respect to the 
distribution (\ref{equ:Pall}), identifying
$a_i=(1-f_i)s^\mathrm{A}_i s^\mathrm{B}_i$.

\subsection{Small-$f$ limit}

Correlation functions between the $a_i$ and the $f_i$ in the 
long-time limit may be calculated in an expansion about $f=0$.   
The idea is simply to collect together configurations $\bm{f}$ 
with exactly $0,1,2,\dots$ frozen spins.  Formally,
one expands $P_\mathrm{f}(\bm{f})$ in Eq.~(\ref{equ:Pall}) 
over $\ee^{-\mu}$, so that
for any observable $Y$,
\begin{equation}
\langle Y \rangle = \frac1\Zf \left[ \langle Y \rangle_0 + \ee^{-\mu} 
\sum_j \langle Y \rangle_j + \ee^{-2\mu} \sum_{j<k} \langle Y \rangle_{jk} 
+ \dots \right]
\label{equ:fexp-general}
\end{equation}
where $\langle Y \rangle_{jkl\dots}$ is an average over configurations 
$A$ and $B$, given that spins $jkl\dots$ are frozen and all other spins
are unfrozen.
(The average $\langle Y \rangle_0$ is taken without any frozen spins, 
so that $A$ and $B$ are independent
configurations from the Boltzmann distribution and averages factorise 
as $\langle f_1(\bm{s}^A) f_2(\bm{s}^B)\rangle_0 = \langle f_1(\bm{s}^A) 
\rangle_0 \langle f_2(\bm{s}^B) \rangle_0$.)  

To make progress, the key step is to write the individual expectation 
values in Eq.~(\ref{equ:fexp-general}) as correlations
with respect to the distribution $P_1(\bm{s}^A) P_1(\bm{s}^B)$, in 
which case configurations $A$ and $B$ are independently
chosen configurations from thermal equilibrium (at $f=0$).  We 
write $\langle \cdots \rangle_0$ for averages
with respect to this distribution.  Using Eq.~(\ref{equ:Pall}), the 
general result is
\begin{equation}
\langle Y \rangle_{k_1k_2\dots k_p} = \left\langle  Y|_{\bm f} \cdot 
\frac{2^{-p} \prod_{r=1}^p (1+s^A_{k_r} s^B_{k_r})}
{Z_B(\bm{f},s^A)}
\right\rangle_0 ,
\label{equ:ave-fixed-f}
\end{equation}
where the notation $Y|_{\bm{f}}$ indicates that any $f$-dependence 
of $Y|_{\bm f}$ has been accounted for by substituting
the specific set of frozen spins $k_1\dots k_p$.  For example 
if $Y=a_i f_j$ as in the main text
then $Y|_{\bm{f}} = a_i \sum_{r=1}^p \delta_{j,k_r}$ since $Y=0$ unless $f_j=1$.
To obtain (\ref{equ:ave-fixed-f}) we used $\delta_{ss'}=\frac12(1+ss')$ 
for Ising spins $s$, $s'$, and we also have that
\begin{equation}
Z_B(\bm{f},s^A) = 2^{-p} \left\langle \prod_{r=1}^p (1+s^A_{k_r} s_{k_r}) 
\right\rangle_s,
\end{equation}
where the average is taken 
over the spins $s$, i.e.,
$\langle \cdots \rangle_s = \sum_{\bm{s}} (\cdots) \ee^{-E(\bm{s})/T}/Z_1$.  

For example, Eq.~(\ref{equ:fexp-general}) can be used to calculate 
$q=\langle a_i \rangle$, leading to
\begin{multline}
q = \langle s_i \rangle_0^2 +  \ee^{-\mu} \left[ 
\sum_{j(\neq i)}  \left\langle s_i^A s_i^B \frac{1+s_j^A s_j^B}{1
+s_j^A \langle s_j\rangle_0}  \right\rangle_0 -V\langle s_i
\rangle_0^2 \right] \\ + O(\ee^{-2\mu}),
\label{equ:aiflin}
\end{multline}
where we used $\Zf=(1+\ee^{-\mu})^V$ and we assumed that averages are translationally invariant, 
$\langle s_i\rangle_0 = \langle s_j\rangle_0$ etc.

If the system has inversion symmetry so that $\langle s_i\rangle_0=0$ 
then the denominator in Eq.~(\ref{equ:aiflin}) is trivial
and the average may be evaluated directly.  Physically, the denominator
 accounts for the fact that the autocorrelation
$\langle a_i\rangle_{\bm f}$ depends on the state of $s_i$ in the 
reference configuration $\bm{s}^A$, and the different
values of $s_i^A$ may not be equally likely.  In the symmetric case, 
both values of
$s_i^A$ are equally likely so the denominator has a trivial value.

To make progress with the general case, note that $s_i=\pm1$ which 
means that that for any function $F(s)$ then
$F(s)=\frac12(1+s)f(1) + \frac12(1-s)f(-1)$ and hence (for any $x$), 
$1/(1+xs_i) = (1-xs_i)/(1-x^2)$.   The result is that
\begin{equation}
q = \langle s_i \rangle_0^2 + f\sum_{j(\neq i)} \frac{\langle \delta 
s_i \delta s_j \rangle_0^2}{1-\langle s_j\rangle_0^2}
 + O(f^2).
\label{equ:qflin}
\end{equation}
One may similarly show that
\begin{equation}
X_{ij} = f \frac{\langle \delta s_i \delta s_j \rangle_0^2}{1-\langle 
s_j\rangle_0^2}
 + O(f^2),
\label{equ:Xflin}
\end{equation}
and summing over $j$ and integrating with respect to $f$ yields 
Eq.~(\ref{equ:qflin}), via Eq.~(\ref{equ:xchi}). 

Physically, $X_{ij}/f$ is the change in the autocorrelation 
$\langle a_i\rangle$ if one restricts to an ensemble where $f_j=1$.
If the spin system has two-point correlations of range $\xi$ 
then  Eq.~(\ref{equ:Xflin}) shows that pinned
spins influence mobile spins over a range of at least $\xi$, 
resulting in an $O(f)$ contribution to $q$.  Of course,
if two-spin correlation functions dominate the physics then the 
pinning procedure is redundant since the correlations may already
be measured through the spin-spin correlation function.  For the 
spin models we consider in this paper, multi-spin correlations
dominate the physics, and we find that $X_{ij}$ is a useful way to 
reveal the relevant correlation lengths without requiring explicit 
measurement of multi-spin correlations.

\end{appendix}

\bibliography{glass}

18 gid=1552001941
18 uid=1632763755
20 ctime=1323256635
20 atime=1323256638
38 LIBARCHIVE.creationtime=1323255414
24 SCHILY.dev=234881026
23 SCHILY.ino=12200485
18 SCHILY.nlink=1


\begin{thebibliography}{44}
\expandafter\ifx\csname natexlab\endcsname\relax\def\natexlab#1{#1}\fi
\expandafter\ifx\csname bibnamefont\endcsname\relax
  \def\bibnamefont#1{#1}\fi
\expandafter\ifx\csname bibfnamefont\endcsname\relax
  \def\bibfnamefont#1{#1}\fi
\expandafter\ifx\csname citenamefont\endcsname\relax
  \def\citenamefont#1{#1}\fi
\expandafter\ifx\csname url\endcsname\relax
  \def\url#1{\texttt{#1}}\fi
\expandafter\ifx\csname urlprefix\endcsname\relax\def\urlprefix{URL }\fi
\providecommand{\bibinfo}[2]{#2}
\providecommand{\eprint}[2][]{\url{#2}}

\bibitem[{\citenamefont{Ediger et~al.}(1996)\citenamefont{Ediger, Angell, and
  Nagel}}]{ediger96}
\bibinfo{author}{\bibfnamefont{M.~D.} \bibnamefont{Ediger}},
  \bibinfo{author}{\bibfnamefont{C.~A.} \bibnamefont{Angell}},
  \bibnamefont{and} \bibinfo{author}{\bibfnamefont{S.~R.} \bibnamefont{Nagel}},
  \bibinfo{journal}{J. Phys. Chem.} \textbf{\bibinfo{volume}{100}},
  \bibinfo{pages}{13200} (\bibinfo{year}{1996}).

\bibitem[{\citenamefont{Debenedetti and Stillinger}(2001)}]{deb-still01}
\bibinfo{author}{\bibfnamefont{P.~G.} \bibnamefont{Debenedetti}}
  \bibnamefont{and} \bibinfo{author}{\bibfnamefont{F.~H.}
  \bibnamefont{Stillinger}}, \bibinfo{journal}{Nature}
  \textbf{\bibinfo{volume}{410}}, \bibinfo{pages}{259} (\bibinfo{year}{2001}).

\bibitem[{\citenamefont{Berthier and Biroli}(2011)}]{rmp_bb}
\bibinfo{author}{\bibfnamefont{L.}~\bibnamefont{Berthier}} \bibnamefont{and}
  \bibinfo{author}{\bibfnamefont{G.}~\bibnamefont{Biroli}},
  \bibinfo{journal}{Rev. Mod. Phys.} \textbf{\bibinfo{volume}{83}},
  \bibinfo{pages}{587} (\bibinfo{year}{2011}).

\bibitem[{\citenamefont{Bouchaud and Biroli}(2004)}]{bouchaud2004}
\bibinfo{author}{\bibfnamefont{J.-P.} \bibnamefont{Bouchaud}} \bibnamefont{and}
  \bibinfo{author}{\bibfnamefont{G.}~\bibnamefont{Biroli}},
  \bibinfo{journal}{J. Chem. Phys.} \textbf{\bibinfo{volume}{121}},
  \bibinfo{pages}{7347} (\bibinfo{year}{2004}).

\bibitem[{\citenamefont{Montanari and Semerjian}(2006)}]{montanari2006}
\bibinfo{author}{\bibfnamefont{A.}~\bibnamefont{Montanari}} \bibnamefont{and}
  \bibinfo{author}{\bibfnamefont{G.}~\bibnamefont{Semerjian}},
  \bibinfo{journal}{J. Stat. Phys.} \textbf{\bibinfo{volume}{125}},
  \bibinfo{pages}{22} (\bibinfo{year}{2006}).

\bibitem[{\citenamefont{Kurchan and Levine}(2011)}]{kurchan2011}
\bibinfo{author}{\bibfnamefont{J.}~\bibnamefont{Kurchan}} \bibnamefont{and}
  \bibinfo{author}{\bibfnamefont{D.}~\bibnamefont{Levine}},
  \bibinfo{journal}{J. Phys. A} \textbf{\bibinfo{volume}{44}},
  \bibinfo{pages}{035001} (\bibinfo{year}{2011}).

\bibitem[{\citenamefont{Cavagna et~al.}(2007)\citenamefont{Cavagna, Grigera,
  and Verrocchio}}]{cavagna2007}
\bibinfo{author}{\bibfnamefont{A.}~\bibnamefont{Cavagna}},
  \bibinfo{author}{\bibfnamefont{T.~S.} \bibnamefont{Grigera}},
  \bibnamefont{and}
  \bibinfo{author}{\bibfnamefont{P.}~\bibnamefont{Verrocchio}},
  \bibinfo{journal}{Phys. Rev. Lett.} \textbf{\bibinfo{volume}{98}},
  \bibinfo{pages}{187801} (\bibinfo{year}{2007}).

\bibitem[{\citenamefont{Biroli et~al.}(2008)\citenamefont{Biroli, Bouchaud,
  Cavagna, Grigera, and Verrocchio}}]{cavagna2008}
\bibinfo{author}{\bibfnamefont{G.}~\bibnamefont{Biroli}},
  \bibinfo{author}{\bibfnamefont{J.-P.} \bibnamefont{Bouchaud}},
  \bibinfo{author}{\bibfnamefont{A.}~\bibnamefont{Cavagna}},
  \bibinfo{author}{\bibfnamefont{T.~S.} \bibnamefont{Grigera}},
  \bibnamefont{and}
  \bibinfo{author}{\bibfnamefont{P.}~\bibnamefont{Verrocchio}},
  \bibinfo{journal}{Nature Phys.} \textbf{\bibinfo{volume}{4}},
  \bibinfo{pages}{771} (\bibinfo{year}{2008}).

\bibitem[{\citenamefont{Kob. et~al.}(2011)\citenamefont{Kob., Roldan-Vargas,
  and Berthier}}]{kob2011}
\bibinfo{author}{\bibfnamefont{W.}~\bibnamefont{Kob.}},
  \bibinfo{author}{\bibfnamefont{S.}~\bibnamefont{Roldan-Vargas}},
  \bibnamefont{and} \bibinfo{author}{\bibfnamefont{L.}~\bibnamefont{Berthier}},
  \bibinfo{journal}{Nature Phys. (in press)}  (\bibinfo{year}{2011}).

\bibitem[{boo()}]{bookdh}
\bibinfo{note}{{\it Dynamical heterogeneities in glasses, colloids, and
  granular media}, Eds.: L. Berthier, G. Biroli, J.-P. Bouchaud, L. Cipelletti,
  and W. van Saarloos, (Oxford University Press, Oxford, 2011).}

\bibitem[{ber()}]{berthier-kob-pts}
\bibinfo{note}{L. Berthier and W. Kob, Phys. Rev. E (in press);
  arXiv:1105.6203}.

\bibitem[{cam()}]{cammarota2011}
\bibinfo{note}{C. Cammarota and G. Biroli, arXiv:1106.5513}.

\bibitem[{\citenamefont{Lipowski}(1997)}]{lipowski1997}
\bibinfo{author}{\bibfnamefont{A.}~\bibnamefont{Lipowski}},
  \bibinfo{journal}{J. Phys. A} \textbf{\bibinfo{volume}{30}},
  \bibinfo{pages}{7365} (\bibinfo{year}{1997}).

\bibitem[{\citenamefont{Newman and Moore}(1999)}]{newman1999}
\bibinfo{author}{\bibfnamefont{M.~E.~J.} \bibnamefont{Newman}}
  \bibnamefont{and} \bibinfo{author}{\bibfnamefont{C.}~\bibnamefont{Moore}},
  \bibinfo{journal}{Phys. Rev. E} \textbf{\bibinfo{volume}{60}},
  \bibinfo{pages}{5068} (\bibinfo{year}{1999}).

\bibitem[{\citenamefont{Garrahan}(2002)}]{garrahan2002-plaq}
\bibinfo{author}{\bibfnamefont{J.~P.} \bibnamefont{Garrahan}},
  \bibinfo{journal}{J. Phys.: Condens. Matt.} \textbf{\bibinfo{volume}{14}},
  \bibinfo{pages}{1571} (\bibinfo{year}{2002}).

\bibitem[{\citenamefont{Jack et~al.}(2005)\citenamefont{Jack, Berthier, and
  Garrahan}}]{plaq-static2005}
\bibinfo{author}{\bibfnamefont{R.~L.} \bibnamefont{Jack}},
  \bibinfo{author}{\bibfnamefont{L.}~\bibnamefont{Berthier}}, \bibnamefont{and}
  \bibinfo{author}{\bibfnamefont{J.~P.} \bibnamefont{Garrahan}},
  \bibinfo{journal}{Phys. Rev. E} \textbf{\bibinfo{volume}{72}},
  \bibinfo{pages}{016103} (\bibinfo{year}{2005}).

\bibitem[{\citenamefont{Jack and Garrahan}(2005)}]{plaq-caging2005}
\bibinfo{author}{\bibfnamefont{R.~L.} \bibnamefont{Jack}} \bibnamefont{and}
  \bibinfo{author}{\bibfnamefont{J.~P.} \bibnamefont{Garrahan}},
  \bibinfo{journal}{J. Chem. Phys.} \textbf{\bibinfo{volume}{123}},
  \bibinfo{pages}{164508} (\bibinfo{year}{2005}).

\bibitem[{\citenamefont{Buhot and Garrahan}(2002)}]{buhot2002}
\bibinfo{author}{\bibfnamefont{A.}~\bibnamefont{Buhot}} \bibnamefont{and}
  \bibinfo{author}{\bibfnamefont{J.~P.} \bibnamefont{Garrahan}},
  \bibinfo{journal}{Phys. Rev. Lett.} \textbf{\bibinfo{volume}{88}},
  \bibinfo{pages}{225702} (\bibinfo{year}{2002}).

\bibitem[{pla()}]{plaq-fdt2006}
\bibinfo{note}{R.~L.~Jack, L. Berthier and J.~P.~Garrahan, J. Stat. Mech.
  (2006), P12005.}

\bibitem[{\citenamefont{Sausset and Levine}(2011)}]{sausset2011}
\bibinfo{author}{\bibfnamefont{F.}~\bibnamefont{Sausset}} \bibnamefont{and}
  \bibinfo{author}{\bibfnamefont{D.}~\bibnamefont{Levine}},
  \bibinfo{journal}{Phys. Rev. Lett.} \textbf{\bibinfo{volume}{107}},
  \bibinfo{pages}{045501} (\bibinfo{year}{2011}).

\bibitem[{\citenamefont{Kim}(2003)}]{kim2003-pin}
\bibinfo{author}{\bibfnamefont{K.}~\bibnamefont{Kim}},
  \bibinfo{journal}{Europhys. Lett.} \textbf{\bibinfo{volume}{61}},
  \bibinfo{pages}{790} (\bibinfo{year}{2003}).

\bibitem[{pro()}]{procaccia2011-pin}
\bibinfo{note}{S. Karmarkar and I. Procaccia, arXiv:1105.4053}.

\bibitem[{tar()}]{tarjus2011pin}
\bibinfo{note}{B. Charbonneau, P. Charbonneau, and G. Tarjus, arXiv:1108.2494}.

\bibitem[{\citenamefont{Kirkpatrick and Thirumalai}(1987)}]{kirk1987}
\bibinfo{author}{\bibfnamefont{T.~R.} \bibnamefont{Kirkpatrick}}
  \bibnamefont{and}
  \bibinfo{author}{\bibfnamefont{D.}~\bibnamefont{Thirumalai}},
  \bibinfo{journal}{Phys. Rev. B} \textbf{\bibinfo{volume}{36}},
  \bibinfo{pages}{5388} (\bibinfo{year}{1987}).

\bibitem[{\citenamefont{Kirkpatrick et~al.}(1989)\citenamefont{Kirkpatrick,
  Thirumalai, and Wolynes}}]{ktw1989}
\bibinfo{author}{\bibfnamefont{T.~R.} \bibnamefont{Kirkpatrick}},
  \bibinfo{author}{\bibfnamefont{D.}~\bibnamefont{Thirumalai}},
  \bibnamefont{and} \bibinfo{author}{\bibfnamefont{P.~G.}
  \bibnamefont{Wolynes}}, \bibinfo{journal}{Phys. Rev. A}
  \textbf{\bibinfo{volume}{40}}, \bibinfo{pages}{1045} (\bibinfo{year}{1989}).

\bibitem[{\citenamefont{Krakoviack}(2011)}]{krakoviack2011-pin}
\bibinfo{author}{\bibfnamefont{V.}~\bibnamefont{Krakoviack}},
  \bibinfo{journal}{Phys. Rev. E} \textbf{\bibinfo{volume}{84}},
  \bibinfo{pages}{050501} (\bibinfo{year}{2011}).

\bibitem[{\citenamefont{Ritort and Sollich}(2003)}]{ritort2003}
\bibinfo{author}{\bibfnamefont{F.}~\bibnamefont{Ritort}} \bibnamefont{and}
  \bibinfo{author}{\bibfnamefont{P.}~\bibnamefont{Sollich}},
  \bibinfo{journal}{Adv. Phys.} \textbf{\bibinfo{volume}{52}},
  \bibinfo{pages}{219} (\bibinfo{year}{2003}).

\bibitem[{\citenamefont{Chandler and Garrahan}(2010)}]{gc2010}
\bibinfo{author}{\bibfnamefont{D.}~\bibnamefont{Chandler}} \bibnamefont{and}
  \bibinfo{author}{\bibfnamefont{J.~P.} \bibnamefont{Garrahan}},
  \bibinfo{journal}{Ann. Rev. Phys. Chem.} \textbf{\bibinfo{volume}{61}},
  \bibinfo{pages}{191} (\bibinfo{year}{2010}).

\bibitem[{\citenamefont{Garrahan and Chandler}(2003)}]{gc2003-pnas}
\bibinfo{author}{\bibfnamefont{J.~P.} \bibnamefont{Garrahan}} \bibnamefont{and}
  \bibinfo{author}{\bibfnamefont{D.}~\bibnamefont{Chandler}},
  \bibinfo{journal}{Proc. Natl. Acad. Sci. USA} \textbf{\bibinfo{volume}{100}},
  \bibinfo{pages}{9710} (\bibinfo{year}{2003}).

\bibitem[{\citenamefont{Berthier and Garrahan}(2005)}]{nef2005}
\bibinfo{author}{\bibfnamefont{L.}~\bibnamefont{Berthier}} \bibnamefont{and}
  \bibinfo{author}{\bibfnamefont{J.~P.} \bibnamefont{Garrahan}},
  \bibinfo{journal}{J. Phys. Chem. B} \textbf{\bibinfo{volume}{109}},
  \bibinfo{pages}{3578} (\bibinfo{year}{2005}).

\bibitem[{\citenamefont{Keys et~al.}(2011)\citenamefont{Keys, Hedges, Garrahan,
  Glotzer, and Chandler}}]{keys2011}
\bibinfo{author}{\bibfnamefont{A.~S.} \bibnamefont{Keys}},
  \bibinfo{author}{\bibfnamefont{L.~O.} \bibnamefont{Hedges}},
  \bibinfo{author}{\bibfnamefont{J.~P.} \bibnamefont{Garrahan}},
  \bibinfo{author}{\bibfnamefont{S.~C.} \bibnamefont{Glotzer}},
  \bibnamefont{and} \bibinfo{author}{\bibfnamefont{D.}~\bibnamefont{Chandler}},
  \bibinfo{journal}{Phys. Rev. X} \textbf{\bibinfo{volume}{1}},
  \bibinfo{pages}{021013} (\bibinfo{year}{2011}).

\bibitem[{\citenamefont{Sasa}(2010)}]{sasa2010}
\bibinfo{author}{\bibfnamefont{S.}~\bibnamefont{Sasa}}, \bibinfo{journal}{J.
  Phys. A} \textbf{\bibinfo{volume}{43}}, \bibinfo{pages}{465002}
  (\bibinfo{year}{2010}).

\bibitem[{\citenamefont{Kim et~al.}(2011)\citenamefont{Kim, Miyazaki, and
  Saito}}]{kim2011-pin}
\bibinfo{author}{\bibfnamefont{K.}~\bibnamefont{Kim}},
  \bibinfo{author}{\bibfnamefont{K.}~\bibnamefont{Miyazaki}}, \bibnamefont{and}
  \bibinfo{author}{\bibfnamefont{S.}~\bibnamefont{Saito}}, \bibinfo{journal}{J.
  Phys.: Condens. Matt.} \textbf{\bibinfo{volume}{23}}, \bibinfo{pages}{234123}
  (\bibinfo{year}{2011}).

\bibitem[{\citenamefont{Berthier
  et~al.}(2007{\natexlab{a}})\citenamefont{Berthier, Biroli, Bouchaud, Kob,
  Miyazaki, and Reichman}}]{berthier2007-jcp1}
\bibinfo{author}{\bibfnamefont{L.}~\bibnamefont{Berthier}},
  \bibinfo{author}{\bibfnamefont{G.}~\bibnamefont{Biroli}},
  \bibinfo{author}{\bibfnamefont{J.-P.} \bibnamefont{Bouchaud}},
  \bibinfo{author}{\bibfnamefont{W.}~\bibnamefont{Kob}},
  \bibinfo{author}{\bibfnamefont{K.}~\bibnamefont{Miyazaki}}, \bibnamefont{and}
  \bibinfo{author}{\bibfnamefont{D.~R.} \bibnamefont{Reichman}},
  \bibinfo{journal}{J. Chem. Phys.} \textbf{\bibinfo{volume}{126}},
  \bibinfo{pages}{184503} (\bibinfo{year}{2007}{\natexlab{a}}).

\bibitem[{\citenamefont{Berthier
  et~al.}(2007{\natexlab{b}})\citenamefont{Berthier, Biroli, Bouchaud, Kob,
  Miyazaki, and Reichman}}]{berthier2007-jcp2}
\bibinfo{author}{\bibfnamefont{L.}~\bibnamefont{Berthier}},
  \bibinfo{author}{\bibfnamefont{G.}~\bibnamefont{Biroli}},
  \bibinfo{author}{\bibfnamefont{J.-P.} \bibnamefont{Bouchaud}},
  \bibinfo{author}{\bibfnamefont{W.}~\bibnamefont{Kob}},
  \bibinfo{author}{\bibfnamefont{K.}~\bibnamefont{Miyazaki}}, \bibnamefont{and}
  \bibinfo{author}{\bibfnamefont{D.~R.} \bibnamefont{Reichman}},
  \bibinfo{journal}{J. Chem. Phys.} \textbf{\bibinfo{volume}{126}},
  \bibinfo{pages}{184504} (\bibinfo{year}{2007}{\natexlab{b}}).

\bibitem[{\citenamefont{Berthier et~al.}(2005)\citenamefont{Berthier, Biroli,
  Bouchaud, Cipelletti, El~Masri, L'Hote, Ladieu, and
  Pierno}}]{berthier2005-science}
\bibinfo{author}{\bibfnamefont{L.}~\bibnamefont{Berthier}},
  \bibinfo{author}{\bibfnamefont{G.}~\bibnamefont{Biroli}},
  \bibinfo{author}{\bibfnamefont{J.}~\bibnamefont{Bouchaud}},
  \bibinfo{author}{\bibfnamefont{L.}~\bibnamefont{Cipelletti}},
  \bibinfo{author}{\bibfnamefont{D.}~\bibnamefont{El~Masri}},
  \bibinfo{author}{\bibfnamefont{D.}~\bibnamefont{L'Hote}},
  \bibinfo{author}{\bibfnamefont{F.}~\bibnamefont{Ladieu}}, \bibnamefont{and}
  \bibinfo{author}{\bibfnamefont{M.}~\bibnamefont{Pierno}},
  \bibinfo{journal}{Science} \textbf{\bibinfo{volume}{310}},
  \bibinfo{pages}{1797} (\bibinfo{year}{2005}).

\bibitem[{\citenamefont{Krakoviack}(2010)}]{krak2010}
\bibinfo{author}{\bibfnamefont{V.}~\bibnamefont{Krakoviack}},
  \bibinfo{journal}{Phys. Rev. E} \textbf{\bibinfo{volume}{82}},
  \bibinfo{pages}{061501} (\bibinfo{year}{2010}).

\bibitem[{\citenamefont{Widmer-Cooper et~al.}(2004)\citenamefont{Widmer-Cooper,
  Harrowell, and Fynewever}}]{harrowell2004}
\bibinfo{author}{\bibfnamefont{A.}~\bibnamefont{Widmer-Cooper}},
  \bibinfo{author}{\bibfnamefont{P.}~\bibnamefont{Harrowell}},
  \bibnamefont{and}
  \bibinfo{author}{\bibfnamefont{H.}~\bibnamefont{Fynewever}},
  \bibinfo{journal}{Phys. Rev. Lett.} \textbf{\bibinfo{volume}{93}},
  \bibinfo{pages}{135701} (\bibinfo{year}{2004}).

\bibitem[{\citenamefont{Widmer-Cooper and Harrowell}(2007)}]{harrowell2007}
\bibinfo{author}{\bibfnamefont{A.}~\bibnamefont{Widmer-Cooper}}
  \bibnamefont{and}
  \bibinfo{author}{\bibfnamefont{P.}~\bibnamefont{Harrowell}},
  \bibinfo{journal}{J. Chem. Phys.} \textbf{\bibinfo{volume}{126}},
  \bibinfo{pages}{154503} (\bibinfo{year}{2007}).

\bibitem[{\citenamefont{Cammarota et~al.}(2011)\citenamefont{Cammarota, Biroli,
  Tarzia, and Tarjus}}]{cammarota2011-rg}
\bibinfo{author}{\bibfnamefont{C.}~\bibnamefont{Cammarota}},
  \bibinfo{author}{\bibfnamefont{G.}~\bibnamefont{Biroli}},
  \bibinfo{author}{\bibfnamefont{M.}~\bibnamefont{Tarzia}}, \bibnamefont{and}
  \bibinfo{author}{\bibfnamefont{G.}~\bibnamefont{Tarjus}},
  \bibinfo{journal}{Phys. Rev. Lett.} \textbf{\bibinfo{volume}{106}},
  \bibinfo{pages}{115705} (\bibinfo{year}{2011}).

\bibitem[{\citenamefont{Castellana}(2011)}]{castellana2011}
\bibinfo{author}{\bibfnamefont{M.}~\bibnamefont{Castellana}},
  \bibinfo{journal}{EPL} \textbf{\bibinfo{volume}{95}}, \bibinfo{pages}{47014}
  (\bibinfo{year}{2011}).

\bibitem[{moo()}]{moore2011-cm}
\bibinfo{note}{J. Yeo and M.~A.~Moore, arXiv:1111.3105}.

\bibitem[{ang()}]{angelini-cm}
\bibinfo{note}{M.~C.~Angelini, G. Parisi and F. Ricci-Tersenghi,
  arXiv:1111.6869}.

\bibitem[{\citenamefont{Berthier and Jack}(2007)}]{berthier-jack2007}
\bibinfo{author}{\bibfnamefont{L.}~\bibnamefont{Berthier}} \bibnamefont{and}
  \bibinfo{author}{\bibfnamefont{R.~L.} \bibnamefont{Jack}},
  \bibinfo{journal}{Phys. Rev. E} \textbf{\bibinfo{volume}{76}},
  \bibinfo{pages}{041509} (\bibinfo{year}{2007}).

\end{thebibliography}

\end{document}